\documentclass[12pt,preprint]{aastex}

\slugcomment{Accepted for publication in \apjs}

\shorttitle{The EBEX Instrument}
\shortauthors{EBEX Collaboration}
\usepackage[nolist]{acronym}
\usepackage[margin=0.75in]{geometry}                
\usepackage{graphicx}
\usepackage{subfigure}
\usepackage{amssymb}
\usepackage{xcolor}
\usepackage{fixltx2e}
\usepackage{amsmath}
\usepackage{hyperref}
\usepackage{float}
\usepackage{tikz}
\usepackage{subfiles}
\usepackage{multirow}
\usepackage{multicol}
\usepackage{calrsfs}
\usepackage{morefloats}

\usetikzlibrary{calc,fit,shadows}
\newcommand{\am}[0]{\ensuremath{'}}
\newcommand{\as}[0]{\ensuremath{''\:}}

\newcommand{\PinBRC}{593}
\newcommand{\PoutBRC}{723}
\newcommand{\PperDfMUX}{21}

\newcommand{\TAB}{Table }

\begin{document}


\title{The EBEX Balloon-Borne Experiment - Gondola, Attitude Control, and Control Software}


\author{The EBEX Collaboration: Asad~Aboobaker\altaffilmark{1}, Peter~Ade\altaffilmark{2}, Derek~Araujo\altaffilmark{3, A}, Fran\c{c}ois~Aubin\altaffilmark{4}, Carlo~Baccigalupi\altaffilmark{5}$^{,}$\altaffilmark{15}, Chaoyun~Bao\altaffilmark{4}, Daniel~Chapman\altaffilmark{3}, Joy~Didier\altaffilmark{3, A}, Matt~Dobbs\altaffilmark{6,7}, Will~Grainger\altaffilmark{8}, Shaul~Hanany\altaffilmark{4}, Kyle~Helson\altaffilmark{9}, Seth~Hillbrand\altaffilmark{3}, Johannes~Hubmayr\altaffilmark{10}, Andrew~Jaffe\altaffilmark{11}, Bradley~Johnson\altaffilmark{3}, Terry~Jones\altaffilmark{4}, Jeff~Klein\altaffilmark{4}, Andrei~Korotkov\altaffilmark{9}, Adrian~Lee\altaffilmark{12}, Lorne~Levinson\altaffilmark{13}, Michele~Limon\altaffilmark{3},  Kevin~MacDermid\altaffilmark{6}, Amber~D.~Miller\altaffilmark{3}, Michael~Milligan\altaffilmark{4}, Lorenzo~Moncelsi\altaffilmark{2, 14}, Enzo~Pascale\altaffilmark{2}, Kate~Raach\altaffilmark{4}, Britt~Reichborn-Kjennerud\altaffilmark{3}, Ilan~Sagiv\altaffilmark{13}, \\ Carole~Tucker\altaffilmark{2}, Gregory~S.~Tucker\altaffilmark{9}, Benjamin~Westbrook\altaffilmark{12}, Karl~Young\altaffilmark{4}, Kyle~Zilic\altaffilmark{4}}


\altaffiltext{1}{Jet Propulsion Laboratory, California Institute of Technology, Pasadena, CA 91109} 
\altaffiltext{2}{School of Physics and Astronomy, Cardiff University, Cardiff, CF24 3AA, United Kingdom} 
\altaffiltext{3}{Physics Department, Columbia University, New York, NY 10027} 
\altaffiltext{4}{University of Minnesota School of Physics and Astronomy, Minneapolis, MN 55455} 
\altaffiltext{5}{Astrophysics Sector, SISSA, Trieste, 34014, Italy} 
\altaffiltext{6}{McGill University, Montr´eal, Quebec, H3A 2T8, Canada} 
\altaffiltext{7}{Canadian Institute for Advanced Research, Toronto, ON, M5G1Z8, Canada} 
\altaffiltext{8}{Rutherford Appleton Lab, Harwell Oxford, OX11 0QX} 
\altaffiltext{9}{Brown University, Providence, RI 02912} 
\altaffiltext{10}{National Institute of Standards and Technology, Boulder, CO 80305} 
\altaffiltext{11}{Department of Physics, Imperial College, London, SW7 2AZ, United Kingdom} 
\altaffiltext{12}{Department of Physics, University of California, Berkeley, Berkeley, CA 94720} 
\altaffiltext{13}{Weizmann Institute of Science, Rehovot 76100, Israel} 
\altaffiltext{14}{California Institute of Technology, Pasadena, CA 91125} 
\altaffiltext{15}{INFN, Sezione di Trieste, Via Valerio, 2, I-34127, Trieste, Italy} 
\altaffiltext{A}{Corresponding Authors: Joy Didier (didier.joy@gmail.com), Derek Araujo (derek@phys.columbia.edu)}


\begin{abstract}
The E and B Experiment (EBEX) was a long-duration balloon-borne instrument designed to measure the polarization of the cosmic 
microwave background (CMB) radiation. 
EBEX was the first balloon-borne instrument to implement a kilo-pixel array of transition edge sensor (TES) bolometric detectors and the 
first CMB experiment to use the digital version of the frequency domain multiplexing system for readout of the TES array. 
The scan strategy relied on 40~s peak-to-peak constant velocity azimuthal scans. 
We discuss the unique demands on the design and operation of the payload that resulted from
these new technologies and the scan strategy. We describe the solutions implemented including   
the development of a power system designed to provide a total of at least 2.3~kW, 
a cooling system to dissipate  590~W consumed by 
the detectors' readout system, software to manage and handle the data of the kilo-pixel array, and 
specialized attitude reconstruction software. We present flight performance data showing faultless management of the TES array, adequate powering and cooling of the readout electronics, 
and constraint of attitude reconstruction errors such that the spurious {\it B}-modes they induced were less than 10\% 
of CMB {\it B}-mode power spectrum with $r=0.05$. 
\end{abstract}




\maketitle

\section{Introduction}
\label{sec:introduction}

Measurements of the \ac{CMB} have provided a
wealth of information about the physical mechanisms responsible for
the evolution of the Universe.
In recent years experimental \ac{CMB} efforts have focused on polarization measurements. 
The polarization signals consist of two distinct patterns: {\it E}-modes and {\it B}-modes \citep{zaldarriaga97}. 
The level and specific shape of the angular power spectrum of CMB {\it E}-mode
polarization can be predicted given the measured intensity anisotropy. 
Lensing of {\it E}-modes by the large scale structure of the Universe produces
cosmological {\it B}-modes at small angular scales. An inflationary phase at sufficiently 
high energy scales near the big bang is predicted to leave another detectable {\it B}-mode signature at large 
and intermediate angular scales \citep{baumann09}.  

The {\it E}-mode polarization of the cosmic microwave background radiation
was first detected by the \ac{DASI} experiment~\citep{DASI_Emodes}.  Other
experiments soon followed suit~\citep{Scott_2010}.  
The combination of all measurements is in excellent agreement with
predictions. {\it B}-mode polarization from gravitational lensing 
of {\it E}-modes and from Galactic dust emission has also recently been 
detected~\citep{SPT_Bmodes, PolarBear_Bmodes, ACT_Bmodes, bicep2detection, bicep+planck}. 
Intense efforts are ongoing by ground- and balloon-based instruments to
improve the measurements, separate the Galactic from the cosmological signals, and 
identify the inflationary {\it B}-mode signature.


\ac{EBEX} was a balloon-borne CMB polarimeter striving to 
detect or constrain the levels of the inflationary gravitational wave and lensing {\it B}-mode power spectra.  
\ac{EBEX} was also designed to be a technology pathfinder for future \ac{CMB} space missions.
To improve instrument sensitivity, we implemented a kilo-pixel array of \ac{TES} bolometers and planned
for a long duration balloon flight. We included three spectral bands centered on 150, 250, and 410~GHz to 
give sensitivity to both the CMB and the galactic dust foreground.  During first observations after 
reaching float, the instrument operated 504, 342, and 109 detectors at 150, 250, and 410~GHz respectively. 
The combination of 400~deg$^2$ intended survey size and optical system with 0.1~deg resolution gave 
sensitivity to the range $30 < \ell < 1500$ of the angular power spectrum. 
Polarimetry was achieved with a continuously rotating achromatic \ac{HWP}.

Several new technologies have been implemented and tested for the first
time in the \ac{EBEX} instrument.
It was the first balloon-borne experiment to implement a kilo-pixel array of \ac{TES} bolometric detectors.
It was the first to implement a digital frequency domain
multiplexing system to read out the \ac{TES} arrays; this digital 
system was later adopted by a number of ground-based experiments.
The \ac{HWP} was levitated using a \ac{SMB};
this was the first operation of an \ac{SMB} in an astrophysical instrument.


Design and construction of the experiment began in 2005.
A ten-hour engineering flight was launched from Ft.\ Sumner, NM in
2009, and the long-duration science flight was launched from Mc\,Murdo
Station, Antarctica on December 29, 2012.
Because the majority of the 25-day long-duration flight was in January 2013, we refer to this flight as EBEX2013.


This paper is one of a series of papers describing the experiment and its in-flight performance.
\ac{EP1} discusses the telescope and the polarimetric receiver; \ac{EP2} describes the detectors and the readout system; and 
this paper, \ac{EP3}, describes the gondola, the attitude control system, and other support systems.
Several other publications give additional details about the \ac{EBEX} experiment. Some are 
from earlier stages of the program~\citep{Oxley_EBEX2004, Grainger_EBEX2008, Aubin_TESReadout2010, Milligan_Software, ReichbornKjennerud_EBEX2010, klein_HWP, Sagiv_MGrossman2012, Westbrook_Design_Evolution}, and others discuss some
subsystems in more detail~\citep{Dan_thesis, Britt_thesis, Sagiv_thesis, Aubin_thesis, MacDermid_thesis, MacDermid_SPIE2014, Westbrook_thesis, Zilic_thesis, chappy_thesis, chappy_ieee_paper, joy_thesis, joy_ieee_paper, Aubin_MGrossman2015}. 

The requirements from a \ac{CMB} polarimeter using a kilo-pixel array of \ac{TES} bolometers 
with frequency domain multiplexing
placed unique demands on the design and operation of the payload. In Section~\ref{sec:gondola} we 
discuss the overall structure of the payload. We also describe the power system that was sized to provide a total power of at least 2.3~kW and a cooling system implemented to radiate the power 
dissipated inside four detector readout crates. 
Maintaining attitude control to the accuracy required by {\it B}-mode measurements is 
discussed in Section~\ref{sec:attitude_control}, and meeting the flight management challenges imposed 
by a kilo-pixel array operating aboard a balloon-platform is discussed in Section~\ref{sec:flightmanagement}.

\section{Gondola: Mechanical, Power, and Thermal Management}
\label{sec:gondola}

\subsection{Gondola Structure}
\label{gondolastructure}

\begin{figure}[ht!]
\begin{center}
\begin{tabular}{cc}
\includegraphics[height=3.25in]{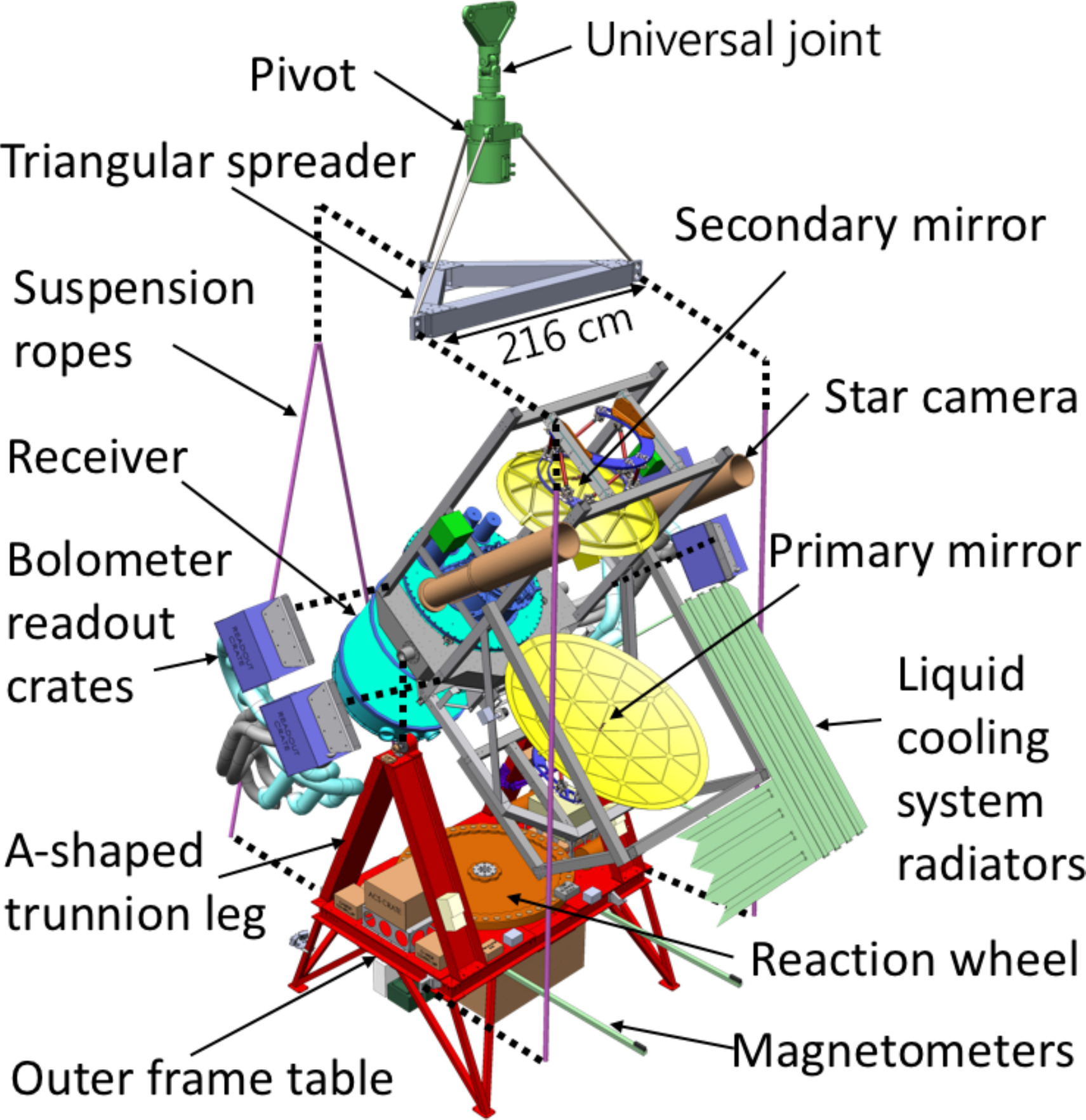} & 
\includegraphics[height=3.25in]{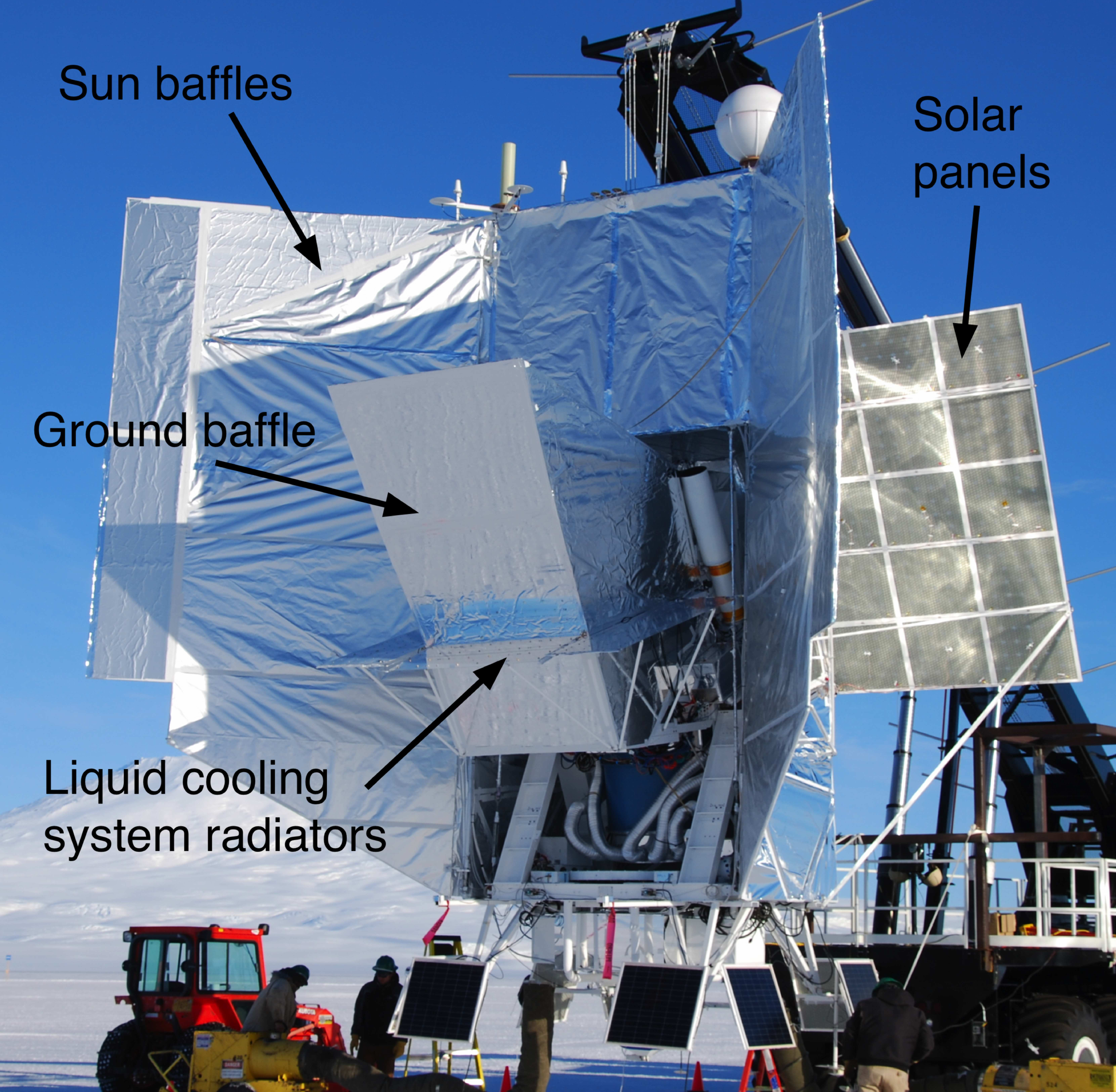}
\end{tabular}
\caption{Left panel: an exploded rendering of the \ac{EBEX} gondola and main components of the instrument. Only the right 
side and half of the front liquid cooling radiators are shown. The other half has been removed for clarity.  Right panel: photograph of the \ac{EBEX} gondola 
before launch.}
\label{figure: gondola}
\end{center}
\end{figure}

We designed the   
\ac{EBEX} gondola, shown in Figure \ref{figure: gondola}, using heritage from past \ac{CMB} balloon payloads 
including the \ac{MAXIMA}~\citep{MAXIMA_instrument} and the \ac{BLAST}~\citep{blast}. The gondola consisted of a rope-suspended outer frame that moved the entire gondola in azimuth and supported an inner frame containing the telescope and 
receiver that moved in elevation.  The science payload weighed 2810~kg, not including NASA's \ac{CSBF} equipment 
and the flight train.

The main structures of both the outer and inner frames were made of 6061 aluminum for its high strength-to-weight ratio 
and ease of fabrication. The 449~kg 
outer frame structure consisted of these main elements: a pivot; three steel turnbuckles connecting the pivot to a 
triangular spreader; four ropes connecting the spreader to a rectangular (2.43~m x 1.68~m) table, made of 
structural I-beams; and A-shaped legs, formed from C-channels, sitting on the two far edges of the table and supporting 
the inner frame; see Figure~\ref{figure: gondola}. The table also held a reaction 
wheel, the flight computers, several coarse attitude sensors, attitude control electronics, and \ac{CSBF} support electronics.

\begin{figure}[ht!]
\begin{center}
\includegraphics[width=0.4\textwidth]{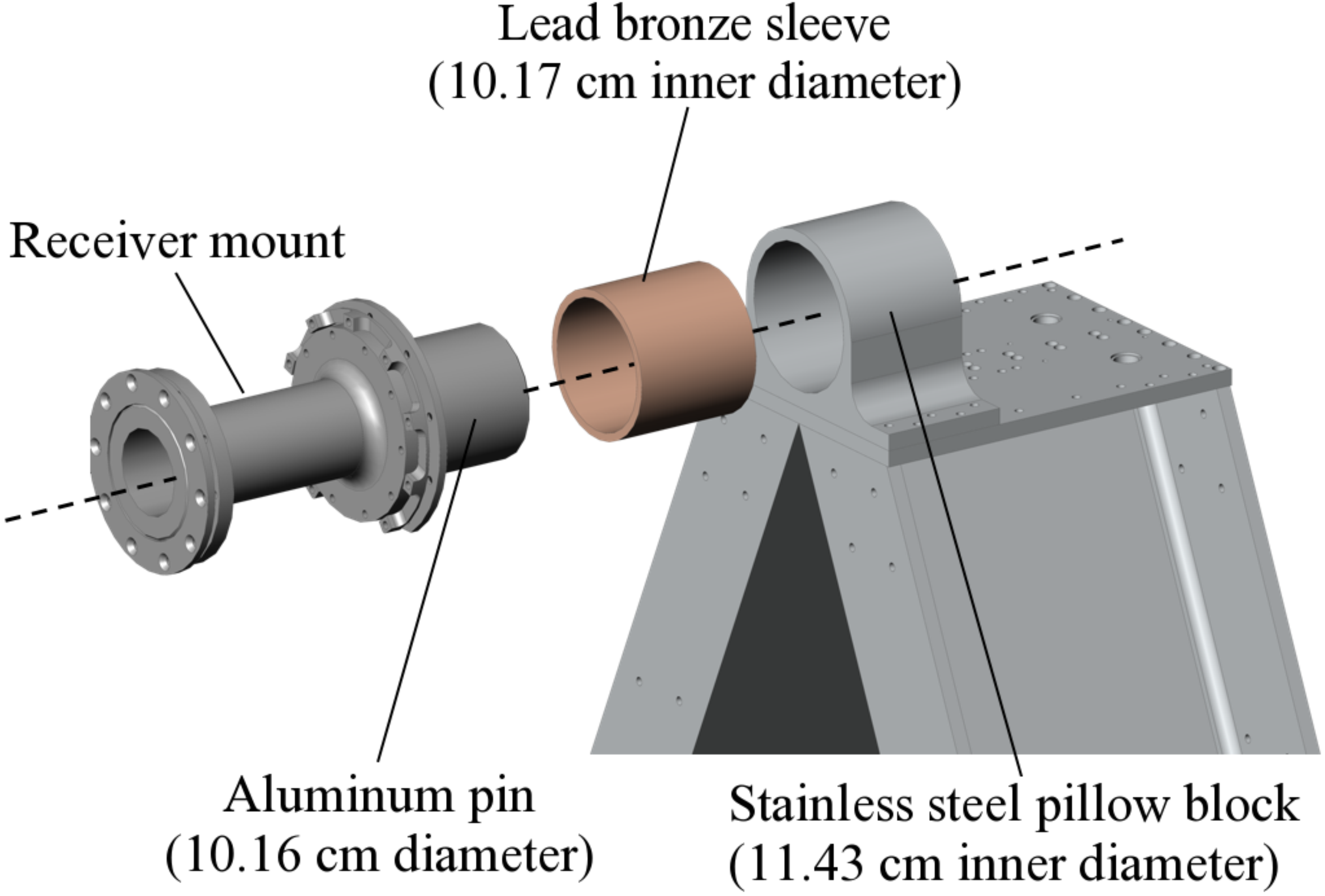}
\caption{Exploded view of one of the two trunnion bearing mounts that supported the inner frame relative to 
the outer frame. The bronze sleeve was press fit into a stainless steel pillow block. The materials and diameters were chosen 
to ensure low friction under the broad range of temperatures encountered during payload ascent and flight. }
\label{figure: pillow_block}
\end{center}
\end{figure}

The 227~kg inner frame consisted of a structure made of box beams that was connected to the receiver and 
that supported the primary and secondary mirrors.  The mirrors were attached to the inner frame by means 
of adjustable hexapods. The hexapods and the rest of the optical systems are described in \ac{EP1}. 
Four bolometer electronic readout crates, two star cameras, 
and two 3-axis gyro boxes were each mounted to the inner frame. The inner frame had a pair of 10.16~cm diameter 
aluminum pins with which the inner frame mounted to trunnion bearings on top of the A-shaped legs; see Figure~\ref{figure: pillow_block}. 
The pins were hard anodized for wear resistance.  Each pin rotated in a 303 stainless steel pillow block mounted atop 
the trunnion legs.  Lead bronze sleeves were pressed into each pillow block in order to reduce friction. The relative dimensions of the pin, 
the lead bronze sleeve, and the pillow block were such that over the expected temperature range the sleeve was always press fit 
in the pillow block while the pin had a diameter clearance that varied between 
0.0122~cm at $-60~^\circ$C and 0.0066~cm at $+40~^\circ$C.  
To minimize stress on the pins due to misalignment, we ensured that the top surfaces of the trunnion legs were aligned such that 
the pins shared a common axis of rotation to within a tolerance of $0.1^\circ$~\citep{Britt_thesis}.

To reduce payload weight, we used Plasma 12-strand ropes\footnote{Puget Sound Rope Corp.} 
made with Spectra polyethylene fiber.\footnote{Honeywell International Inc.}  To our knowledge, this was the first use of Spectra fiber ropes in a stratospheric balloon application.  This raised two concerns.  
First, the strength of the ropes degrades with exposure to ultraviolet (UV) light.  Second, the ropes undergo permanent lengthening, 
or creep, which increases with time, increased load, and temperature.  To address these concerns, we conducted ground and flight 
tests to certify the ropes and quantify the creep and degradation from exposure to solar UV light.  
Laboratory measurements of rope creep over a 16-day period showed that the 508~cm 
ropes lengthened by 0.76~cm 
over a 9-day initial stretching phase, after which the rope length stabilized~\citep{Britt_thesis}.  Because the Antarctic flight ropes 
were pre-stretched while the gondola hung from the ropes during months of pre-flight testing, we assumed that negligible creep 
would occur during flight.  To reduce the degradation in strength anticipated from solar radiation, we shielded the Antarctic flight ropes 
by wrapping them in two layers of aluminized mylar, 
each layer consisting of a 6.35~$\mu$m thick polyester film with a 50~nm~thick layer of vapor-deposited aluminum. Tests of breaking strength conducted after a 
28-hour rope certification flight launched in September 2008 from Ft.~Sumner, NM showed that the shielding provided significant 
protection against the degradation in the breaking strength; the 
bare ropes had an average degradation of nearly 10\%, while the shielded ropes had a degradation of close to 2\% 
(see \TAB\ref{rope_break_tests}).  Assuming an exponential model, consistent with vendor ground testing data~\citep{honeywell_handbook}, 
we calculated the time constant for degradation at float and concluded that it would take 67 days for the breaking strength to 
decrease to the minimum level necessary to support a $10g$ vertical acceleration of the payload, as required by NASA.   

\begin{table}[ht!]
\begin{centering}
Post-Certification Flight Rope Break Test Results \\ 
\begin{tabular}{ll}
\hline \hline
Rope tested & Breaking strength (N) \\ \hline 
Bare rope 1                                & 220,000                          \\ 
Bare rope 2                                & 228,000                           \\ 
Shielded rope 1                              & 246,000                          \\ 
Shielded rope 2                              & 240,000                          \\ 
Reference rope (not flown)                 & 247,000                          \\ \hline 
\end{tabular}
\caption{Results from breaking strength tests of bare and aluminum mylar shielded Spectra fiber ropes flown during a 28-hour certification flight and reference rope (not flown).  The average degradation in breaking strength for the shielded ropes was 4,000~N, 
while the average degradation for the bare ropes was 23,000~N.} 
\label{rope_break_tests}
\end{centering}
\end{table}

We designed lightweight baffles to shield the telescope and receiver from direct illumination by the Sun and Earth. 
Baffles connected to the outer frame gave Sun protection when the azimuth 
of the telescope was within $\pm$60$^\circ$ from anti-Sun and for all Sun elevation angles during any 24~hour period as 
long as the payload was at latitude Southward of 73$^\circ$ South.  Baffles connected to the inner frame provided protection from Earth 
for telescope elevation angles larger than 30$^\circ$, which was the lowest nominal sky observation angle. 

\begin{figure}[ht!]
\begin{center}
\includegraphics[height=2.5in]{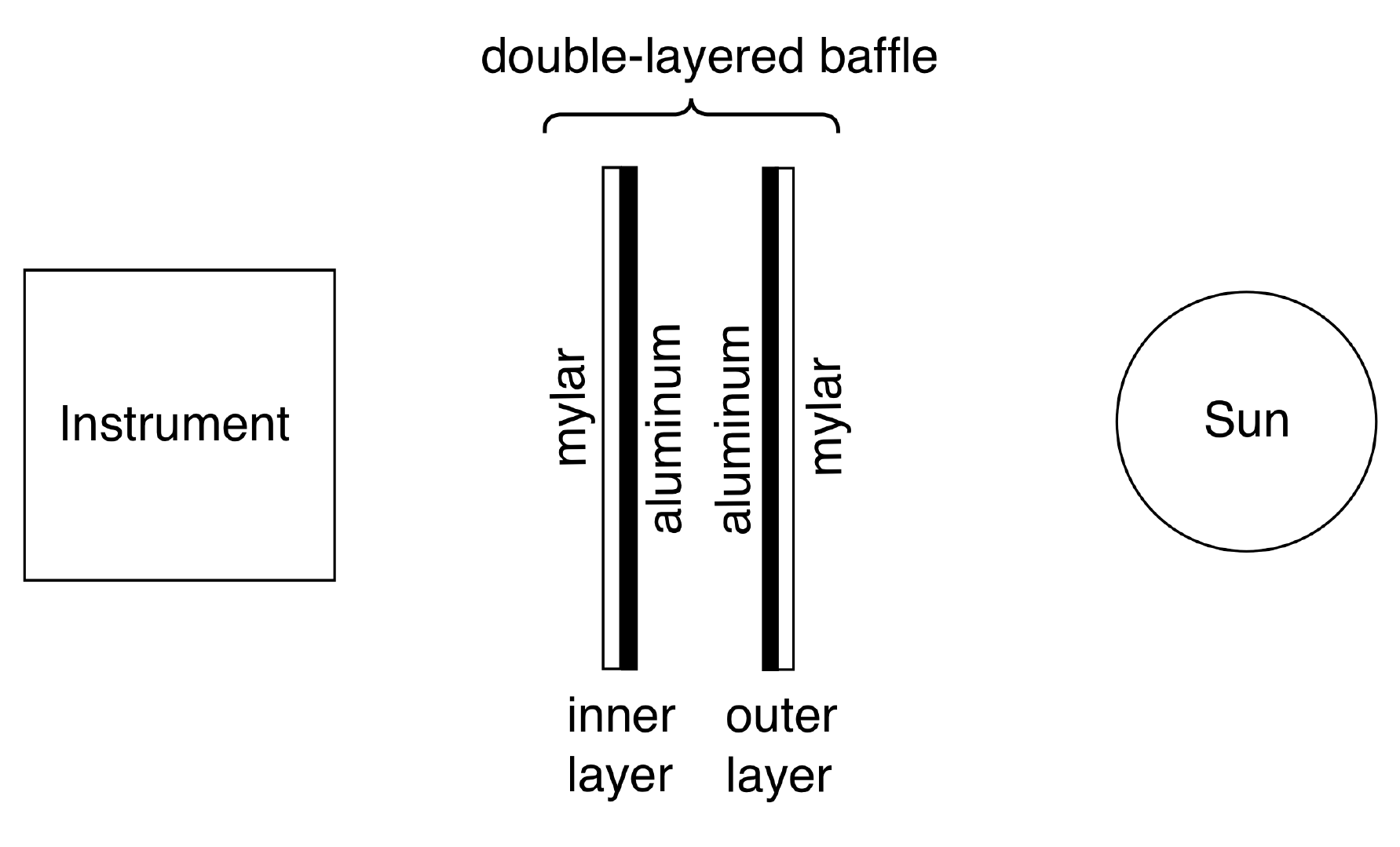}
\caption{Surface orientation of the double-layered aluminized mylar baffles.  Solar radiation is partially reflected and partially absorbed by the 
aluminum layer of the Sun-facing baffle. The mylar layer, which has high infrared emissivity, re-radiates the energy to space. Similar process occurs
for the instrument-facing baffle for scattered solar radiation. }
\label{figure: baffle_drawing}
\end{center}
\end{figure}

Each baffle surface contained two layers of aluminized mylar film\footnote{Lamart Corporation} in a strategy akin to that discussed by~\citet{BLASTPol_thermal}.   
The outer layer used 50~nm thick vapor-deposited aluminum, and the inner layer used 8.9~$\mu$m thick aluminum 
foil bonded to a 50~$\mu$m thick mylar film.  The mylar layers, which have high infrared emissivity and are thus responsible for radiating energy 
to space, were oriented such as to maximize the view factor to the open sky;  see Figure~\ref{figure: baffle_drawing}. 

We fabricated the 90~kg outer baffle structure from welded aluminum tubes designed for sufficient mechanical strength to 
support the triangle spreader and pivot when the suspension ropes were slack.  To minimize weight, we fabricated the 
inner baffle from closed-cell extruded polystyrene foam, and glued the aluminized films to the foam.  Carbon fiber support structures 
would have been lighter but also significantly more expensive.  

\subsection{Gondola Motion Control}
\label{sec:gondolamotion}

\begin{figure}[ht!]
\begin{center}
\includegraphics[height=2.5in]{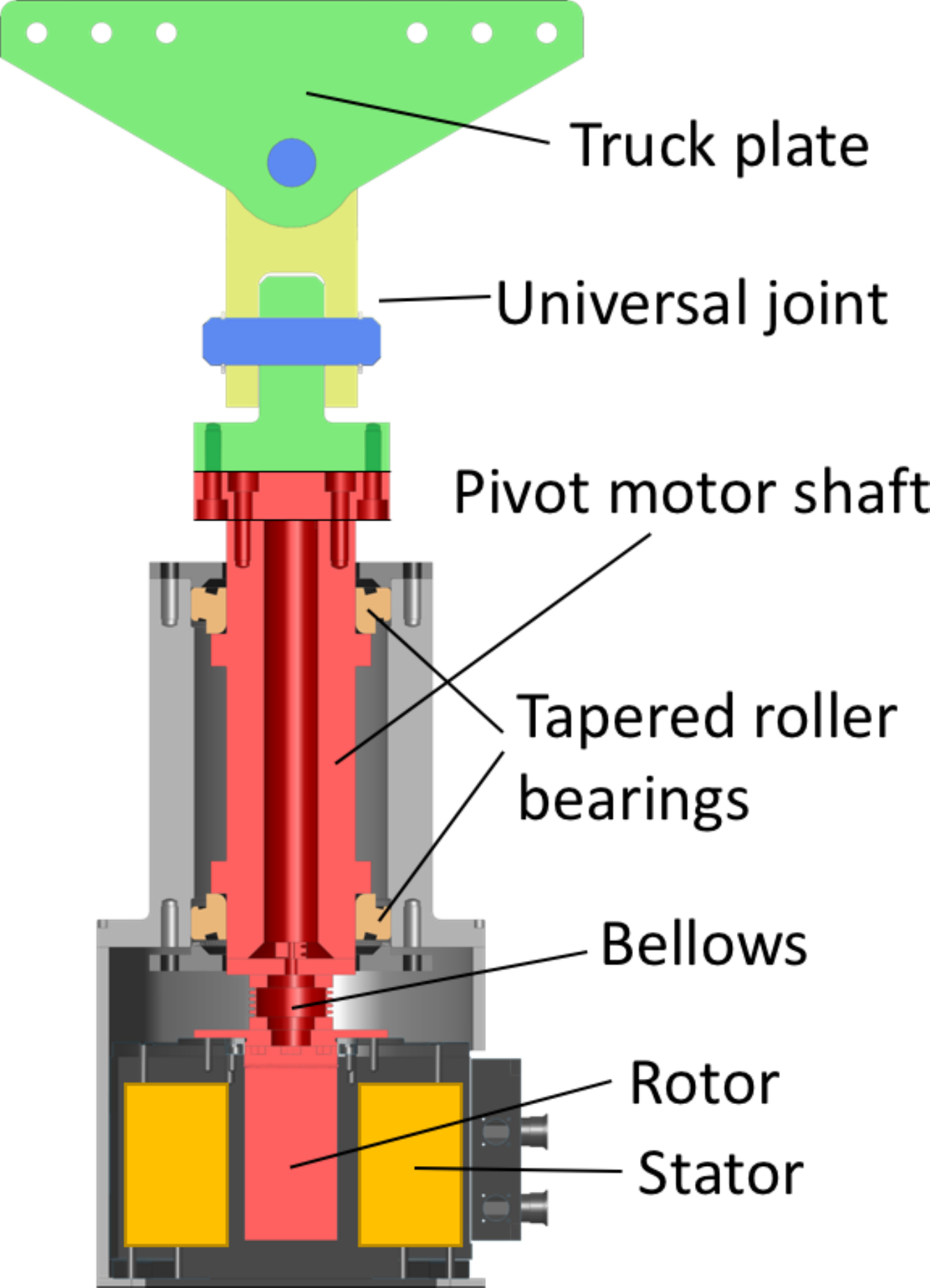}
\caption{The pivot connected the flight train to the gondola and enabled active relative rotation. The flight train was connected to 
a truck plate, a universal joint, and a shaft. The gondola was suspended on the shaft by means of two tapered roller bearings. The rotor of a motor, whose
stator was connected to the gondola, was coupled to the shaft with bellows.}
\label{figure: pivot}
\end{center}
\end{figure}

Azimuth motion control was achieved with an active pivot and a reaction wheel, each of which was driven by a brushless 
DC motor.\footnote{Kollmorgen Model D102M} In nominal 
motion, the pivot motor was intended to torque the entire gondola relative to the flight line and the balloon; the reaction wheel 
was to provide fine tuning of azimuth motion. A detailed description of the azimuth control system can be found in \cite{Britt_thesis}. The reaction wheel had a 
moment of inertia of 50.0~kg~m$^2$, approximately 1.6\% of the total moment of inertia of the gondola. 
The pivot, shown in Figure~\ref{figure: pivot}, consisted of a shaft that was rigidly connected to a universal joint and from there 
to the flight train. The gondola was suspended on the 
shaft by means of two tapered roller bearings. The rotor of the motor was coupled to the shaft with bellows, obviating the need for precise axial alignment.  
All moving parts on the gondola were lubricated with low-temperature 
greases.\footnote{Castrol Braycote 601EF, Dow-Corning Molykote 33 Light, or Mobil Mobilgrease 28}

Because of an error in thermal design for the EBEX2013 Antarctic flight -- see Section~\ref{sec:thermal_management} -- the pivot motor controller 
overheated and shut-down periodically, disabling control of azimuth motion. No such problems occurred during the North American 2009 engineering flight.   
A linear actuator\footnote{SKF USA Inc. Model CARN32} provided elevation motion.  The \ac{EBEX} actuator had a 700~mm stroke and a 
maximum force of 3500~N, and enabled a telescope elevation range of $17^\circ$ to $68^\circ$ (corresponding to an upright inner frame tower).  
We planned to observe the CMB above $30^\circ$ elevation, but maintained the capability for lower angles for observations of planets 
that are occasionally visible from Antarctica.  The actuator was driven by a DC brush motor\footnote{Pittman Motors Model 14207} 
fitted with high-altitude brushes.  To minimize the average force required from the actuator over the course of the flight, the inner frame 
was designed to be balanced with the cryogens in the cryostat half full.  Two motor drives of the same 
model\footnote{ADVANCED Motion Controls Model DR-100RE} controlled the reaction wheel and the pivot motors, and an additional motor drive\footnote{ADVANCED Motion Controls Model 30A8DD} controlled the linear actuator motor \citep{Britt_thesis}.

\begin{figure}[ht!]
\begin{center}
\includegraphics[height=6in]{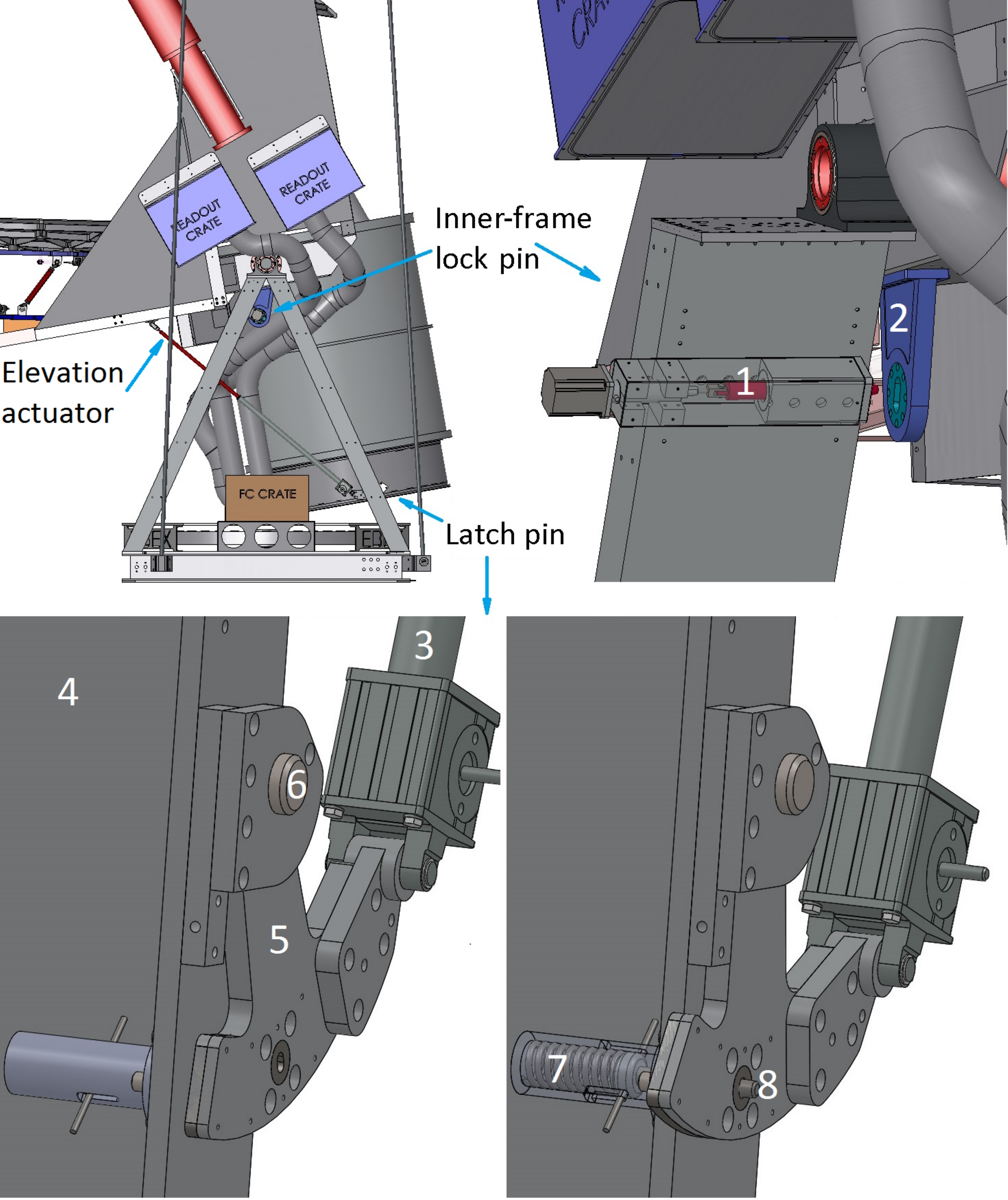}
\caption{The elevation actuator protection mechanism -- top left panel -- had two elements, an `inner frame lock pin' -- top right panel-- and a `latch pin' -- bottom. 
The inner frame lock -- top right -- locked the inner frame to the outer frame by means of a motorized, 2.54~cm diameter, steel pin (1). 
Before launch and before flight termination the motorized pin was commanded to engage into an inner frame bracket (2) that had a slot with tapered walls. 
But even as the inner frame was locked to the outer frame, it could experience elastic deformations exercising excessive axial loads on the
elevation actuator. To protect the actuator, the latch was kept released -- bottom left panel. The end of the actuator (3) was connected to the outer frame (4) 
via a bracket (5) that was allowed to rotate about a pin (6). After launch accelerations subsided, the elevation actuator was commanded to extend, bringing
the hole in the bracket of the latch (5) to alignment with a spring (7) loaded pin (8), thus latching the actuator in place -- bottom right panel. 
The motorized pin of the inner frame lock (1) was then retracted, releasing the inner frame. }
\label{figure: elevation_actuator_lock}
\end{center}
\end{figure}

To protect the elevation linear actuator from excessive loads during launch accelerations, we designed an elevation actuator protection 
mechanism; see Figure~\ref{figure: elevation_actuator_lock}.  The mechanism consisted of an inner frame locking pin driven by a 51~mm stroke 
actuator\footnote{Ultra Motion} and a spring-loaded latch pin attached to the bottom end of the actuator.  
During launch we fixed the position of the inner frame with the locking pin and allowed the bottom end 
of the actuator to undergo limited motion.  After launch, we fixed the bottom actuator end into operating position by engaging a 
spring-loaded pin and retracted the inner frame locking pin.

\subsection{Power}
\label{sec:power}

\begin{figure}[ht!]
\begin{center}
\includegraphics[height=7in]{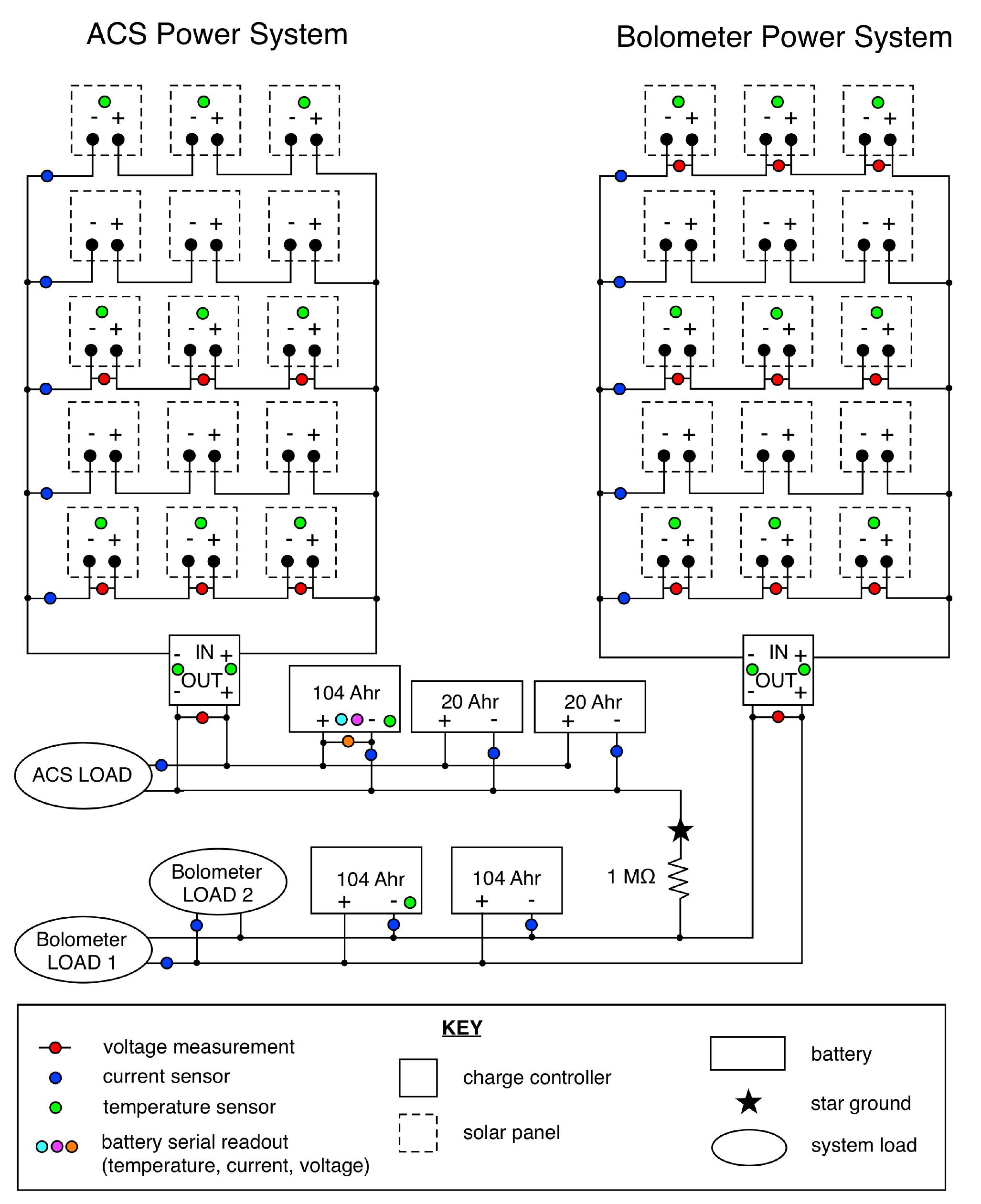}
\caption{Schematic of the \ac{ACS} and bolometer power systems.  The two power systems shared a common electrical ground, 
marked here with a star ($\bigstar$).  Each 28~V power system contained high capacity lithium-ion batteries (with respective 
charge capacities of 144~Ahr and 208~Ahr), charged by 15 lightweight solar panels (each weighing 1~kg and specified to produce 76~W).}
\label{power_schematic}
\end{center}
\end{figure}

Two separate systems generated and supplied power to the detectors and the rest of the electronics; the two systems shared only a single common 
electrical ground point; see Figure~\ref{power_schematic}. 
The bolometer system provided power for detector biasing and readout, cryostat housekeeping 
and refrigerator control, and half wave plate readout and control. The \ac{ACS} system powered the 
flight control computers, attitude control sensors and motors, data transmitters, the liquid cooling system pumps, and heaters for 
the sensors, motors, and batteries.  Total peak power consumption was 1.7~kW as measured on the ground while connected 
to a power supply; see \TAB\ref{power-consumption}.  

\begin{table}[ht!]
\centering
\begin{tabular}{lr}
\multicolumn{2}{c}{ACS Power System}     \\ \hline \hline
Component                            & Power consumption [W] \\ \hline
Flight computer \& data storage            & 189\\ 
Sensors                      &  135    \\ 
Motors               &   66         \\ 
Liquid cooling system pumps     &    40   \\ 
Line of sight video \& data transmitters*  &      181*      \\ 
Heaters (sensor, motor, \& battery)*            &    325*    \\  \hline
Total peak power consumption:         & 936       \\ \hline
          & \multicolumn{1}{l}{}          \\
\multicolumn{2}{c}{Bolometer Power System}    \\ \hline \hline
Component         & Power consumption [W] \\ \hline
DC-DC bias crate      &    137               \\ 
Bolometer readout crates (4 units)          &    586 \\ 
Cryostat housekeeping \& refrigerator control &   46 (7*)         \\ 
Half-wave plate crate            &    32     \\ \hline
Total peak power consumption:          &     801                     \\  \hline
\end{tabular}
\caption{Power consumption by the \ac{EBEX} instrument \ac{ACS} and bolometer power systems as measured on the ground while connected to a power supply.  
Components marked with an asterisk (*) consumed power only intermittently.  Data and video transmitters were active only during the 
first 24 hours. Heaters were active primarily during payload ascent.  Out of the 46~W consumed for cryostat housekeeping and refrigerator control, 7~W were expended only during refrigerator cycling.  Without these intermittent components total power consumption 
was 1224~W.}
\label{power-consumption}
\end{table}

Each solar power system consisted of 15 solar panels\footnote{Suncat Solar LLC} weighing 1~kg each, and covering a 
total area of 8.6~m$^{2}$.  
Each panel was specified to produce 76~W at float assuming orthogonal Sun, zero albedo, and an operating temperature of 110~$^{\circ}$C. 
Under these assumptions the \ac{EBEX} power system was capable of generating 2280~W. 
The power from each array of 15 panels was fed to a charge controller.\footnote{Morningstar Model TS-MPPT-60}  
We replaced the controller heatsink fin with an aluminum block heat sunk directly to the gondola frame.  We chose 
lithium-ion batteries for their high power density; the battery capacity for the detector and \ac{ACS} power domains were 208~Ahr and 144~Ahr, respectively, and the nominal voltage of each system was 28~V.

The batteries included built-in control electronics that provide temperature, voltage, and 
current monitoring. They also included a mechanism to open the 
battery's main switch in case of a detected fault.  We disabled this feature because we found it generated fault detections even under normal conditions. 

To specify the proper solar panel area and battery capacity, we produced a simulation of the power system 
throughout the flight, from the pre-launch hands-off period on the ground through ascent and successive \ac{CMB} and 
calibrator scans at float.  Because the instrument did not complete the scan strategy as planned during 
flight (see Section \ref{sec:thermal_management}), we evaluate the power system performance by comparing the minimum 
battery capacity predicted by the simulation to that reached after ascent, during which the gondola spun in full rotations, as expected
and simulated.  

To ensure that we did not underestimate the required battery capacity, we conservatively assumed a 35\% albedo 
and 110~$^{\circ}$C panel temperature in our simulation. (Solar power production was expected to  
degrade by 0.4\% for every 1~$^{\circ}$C increase in temperature.)   Under these assumptions the simulation predicted a minimum battery 
state of charge of 40\% when the payload reached float altitude. Problems with the battery electronics readout 
prevented reliable monitoring of all but one of the five batteries. That battery's measured state of charge decreased from 94\% to 91\% during 
ascent.  The difference from the simulation is explained by the fact that the actual albedo was probably closer to 100\%. Also, the actual solar 
panel temperature ranged between $-15$ to $+60~^{\circ}$C. During flight the battery reached a minimum state of charge of 65\% 
during the first day of flight, and generally remained between $85-100$\% charged.

\subsection{Thermal Management}
\label{sec:thermal_management}

\subsubsection{Instrument-wide Overview}
\label{sec:thermaloverview}

An engineer working with NASA's \ac{CSBF} performed a thermal analysis of the entire instrument, 
taking into account the radiation from the Sun and Earth at float, the minimum and maximum power dissipated in each component, 
electrical enclosure surface coverings, radiation scattering due to the atmosphere, and the thermal conductivity of the air at float altitude.  
For this analysis the engineer used values for Earth's albedo and long wavelength radiation that are up to 2.3$\sigma$ higher and lower than  
measured means, such that the extreme values would occur in only 2\% of the cases. This analysis yielded
the extreme temperatures that are likely to be encountered by any piece of hardware aboard the payload. 
When the simulation indicated components would exceed their specified operating ranges, we added heaters or improved cooling, as necessary. 
We then used thermal vacuum tests to ensure that each component would operate within its required temperature range. 
In order to minimize infrared absorption by bare aluminum surfaces, the entire gondola frame was white powder coated; 
some electronics enclosures were painted with white Krylon Appliance Epoxy\footnote{Sherwin Williams Company} while for others we used silver Teflon tape\footnote{Sheldahl fluoro-ethylene propylene tape by Multek Corporation}.  

During flight, the pivot motor controller overheated and automatically shut itself down when we attempted to execute the planned scan strategy. 
The overheating resulted from a thermal model error that led to inadequate cooling of the pivot motor controller, which was exposed to direct sunlight.
The error, which went undetected throughout payload development, was introduced when the gondola solid model was imported 
into the thermal model.  
As a result, the azimuth control system did not perform as designed, so (i) we were unable to observe the originally planned sky 
region and (ii) some instrument components operated in an unanticipated radiative environment for a prolonged period.
Nevertheless, when the telescope was pointed away from the Sun, all instrument components except the pivot motor controller 
operated within the expected temperature range, validating all other elements of the thermal design.

\subsubsection{Liquid Cooling System}
\label{sec:thermal_lcs}

\begin{figure}[ht!]
\begin{center}
\includegraphics[height=3in]{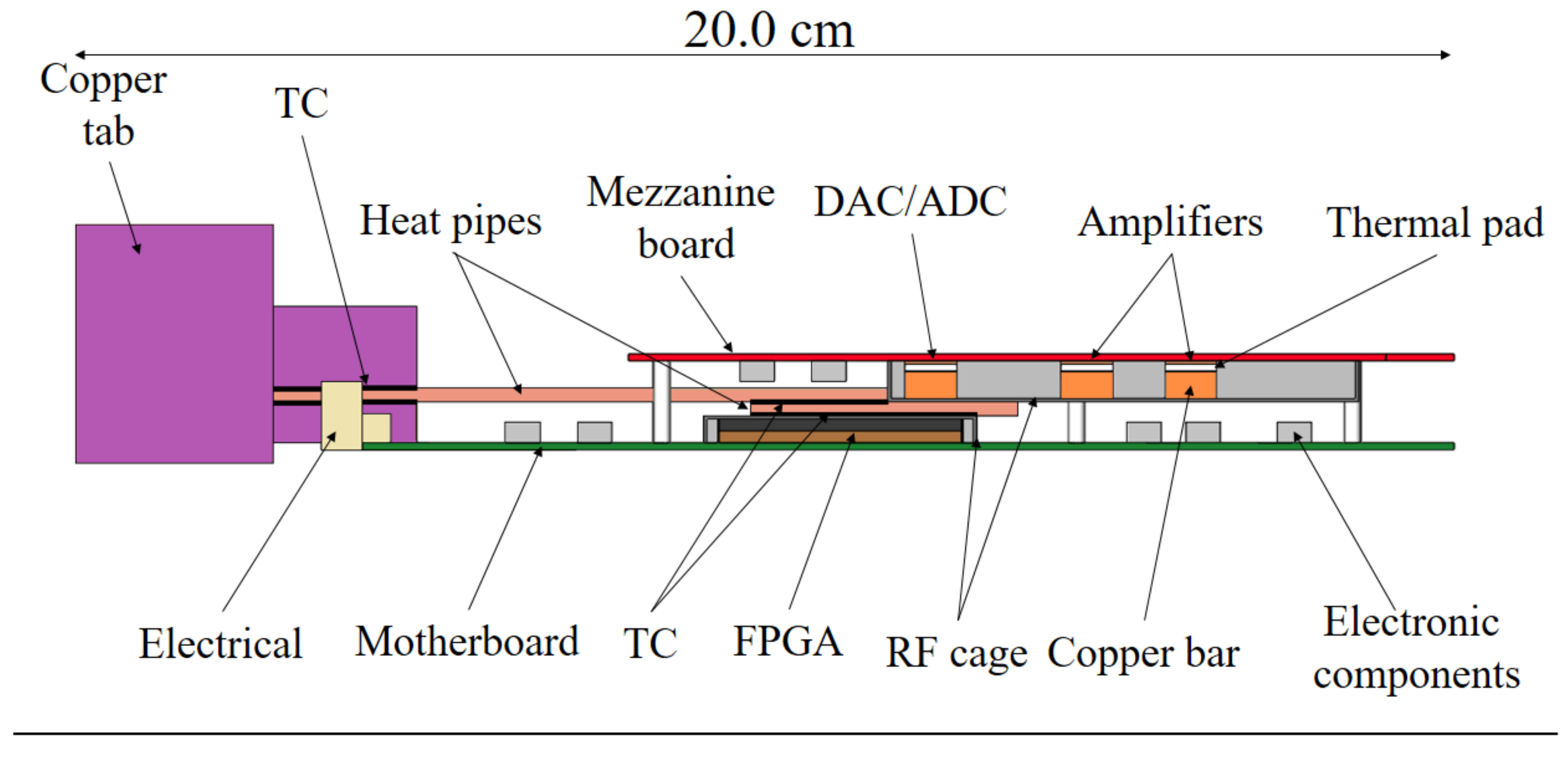}
\vspace{0.3in}
\includegraphics[height=3in]{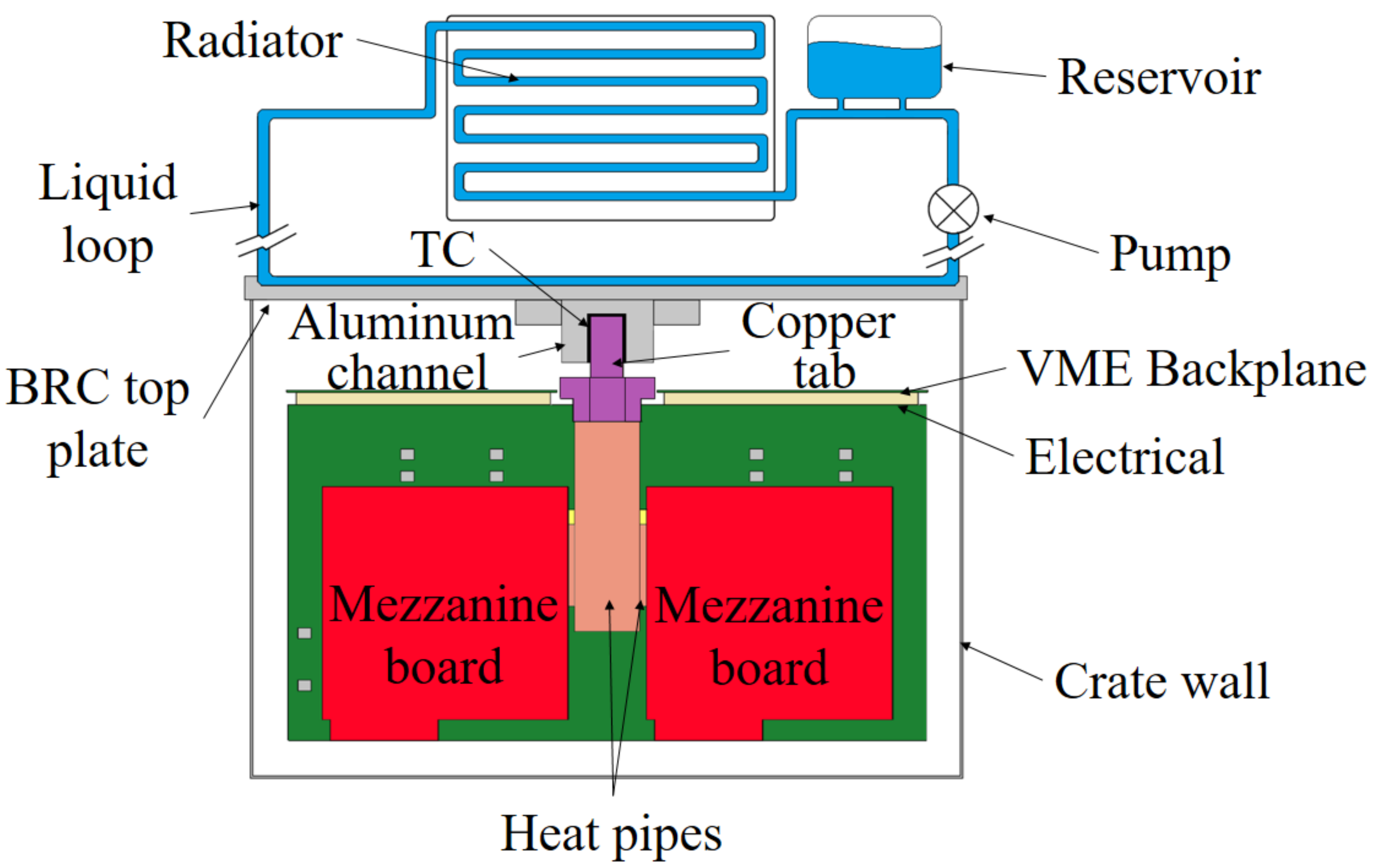}
\caption{Top panel: side view and scale bar of a readout board and its heat dissipating and conducting components. Lower panel: a sketch
of the entire thermal pass including a top view of a readout board, the liquid cooling loop, and radiator panels. Power dissipated by the \ac{FPGA} on the 
motherboard (top panel, brown) was transferred to its RF cage via thermal compound (TC). Power dissipated by the \ac{DAC}/\ac{ADC} 
on the mezzanine boards
was transferred to the RF cage via intermediate copper bars (orange) and thermal interface pads (white) with a thin layer of thermal compound.  
Two flat heat pipes (salmon) that were glued to the RF cages conducted power to a copper tab (purple) that pressed into a channel 
in the BRC top plate (bottom panel, grey). A liquid coolant transferred the power to external radiator panels. }
\label{figure: brc_nanospreaders}
\end{center}
\end{figure}

For detector readout we used 28 \ac{DfMUX} boards distributed among 
four \acp{BRC}. Two \acp{BRC} had 6 and two had 8 boards each. 
The four \acp{BRC} with all the \ac{DfMUX} boards operating, SQUIDs tuned, and all bolometers overbiased drew \PoutBRC~W from 
the power system, of which \PinBRC~W were dissipated inside the \ac{BRC}s and the rest were dissipated in a separate 
power crate that housed the 82\% efficiency DC-DC converters. The \PperDfMUX~W per board were dissipated mostly by a \ac{FPGA} 
on the motherboard and by  \acp{DAC} and \acp{ADC} on each of two mezzanine 
boards that were plugged to the motherboard. Each of these two sources - \ac{FPGA} and \ac{DAC}/\ac{ADC} - was encased 
with an RF enclosure. We designed a thermal cooling system to move the dissipated energy from the hot components to their RF enclosure, 
then from the RF enclosure to the back of the \ac{BRC} and to the ambient environment.  Figure~\ref{figure: brc_nanospreaders}  gives 
the details of the approach we used.  

To provide thermal connection between the \ac{FPGA} and the RF cage, we used a thermally 
conductive compound.\footnote{Arctic Silver 5 by Arctic Silver} For the \ac{DAC}/\ac{ADC} and amplifiers on the mezzanine 
boards we used small copper bars and thermal interface pads\footnote{Part 5519S by 3M} with a thin layer of the thermal compound. 
We used flat heat pipes\footnote{NanoSpreaders made by Celsia Technologies} that were glued with thermal 
adhesive compound\footnote{Arctic Silver Thermal Adhesive by Arctic Silver} to the RF cages to transfer the energy to 
a copper tab at the edge of the board. The copper tab was pressed into a slot on the top of the \ac{BRC} that also 
hosted a liquid cooling loop.\footnote{Lytron CP15 by Lytron} The liquid transferred the heat to radiator panels that had a total 
area of 4.1~m${^2}$. 

During pre-flight vacuum chamber tests we measured a $45~^{\circ}$C temperature difference between the \ac{DfMUX} 
motherboard and the \ac{BRC} wall. The liquid cooling system and the area of the radiators were designed to maintain 
the top of the \ac{BRC} wall below $25~^{\circ}$C so as to 
maintain the \ac{DfMUX} electronics below the most stringent component 
specification limit of $70~^{\circ}$C. The \ac{LCS} consisted of two independent closed loops, each responsible for 
dissipating heat from two \ac{BRC}s.  
In each loop we employed a 20~W, 80~psi differential pressure pump\footnote{Micropump GJ Series} to circulate the
coolant,\footnote{Dynalene HC-40 by Dynalene, Inc.} and a 
reservoir to accommodate pressure variations and prevent bubbles. We bolted multichannel extruded aluminum heat exchangers to the 
radiator panels, and applied a thin layer of thermally conductive silicone  
grease\footnote{Chemplex 1381} between the extrusions and the panels. The total length of tubing for each LCS loop was 13.3~m and the average diameter was 4.75~mm. 
At an average coolant flow of 30~mL~s$^{-1}$ the heat transfer between the bulk of the liquid and the 
heat exchangers, as well as the dynamics of the coolant in the tubes, were 
consistent with the regime of weak turbulence. The pressure gradients across the LCS elements were in a good agreement with the 
turbulent model predictions with a friction factor corresponding to an average wall roughness of 0.1~mm. 
The radiators dissipated the heat to space with an average view factor of 0.52. To minimize solar absorptance and enhance infrared emittance we covered 
the panel surfaces with silver Teflon tape\footnote{Sheldahl fluoro-ethylene propylene tape by Multek Corporation}  with 
solar absorptance $\alpha \le 0.10$ and infrared normal emittance $\epsilon \ge 0.80$.

\begin{figure}[ht!]
\begin{center}
\includegraphics[height=3.5in]{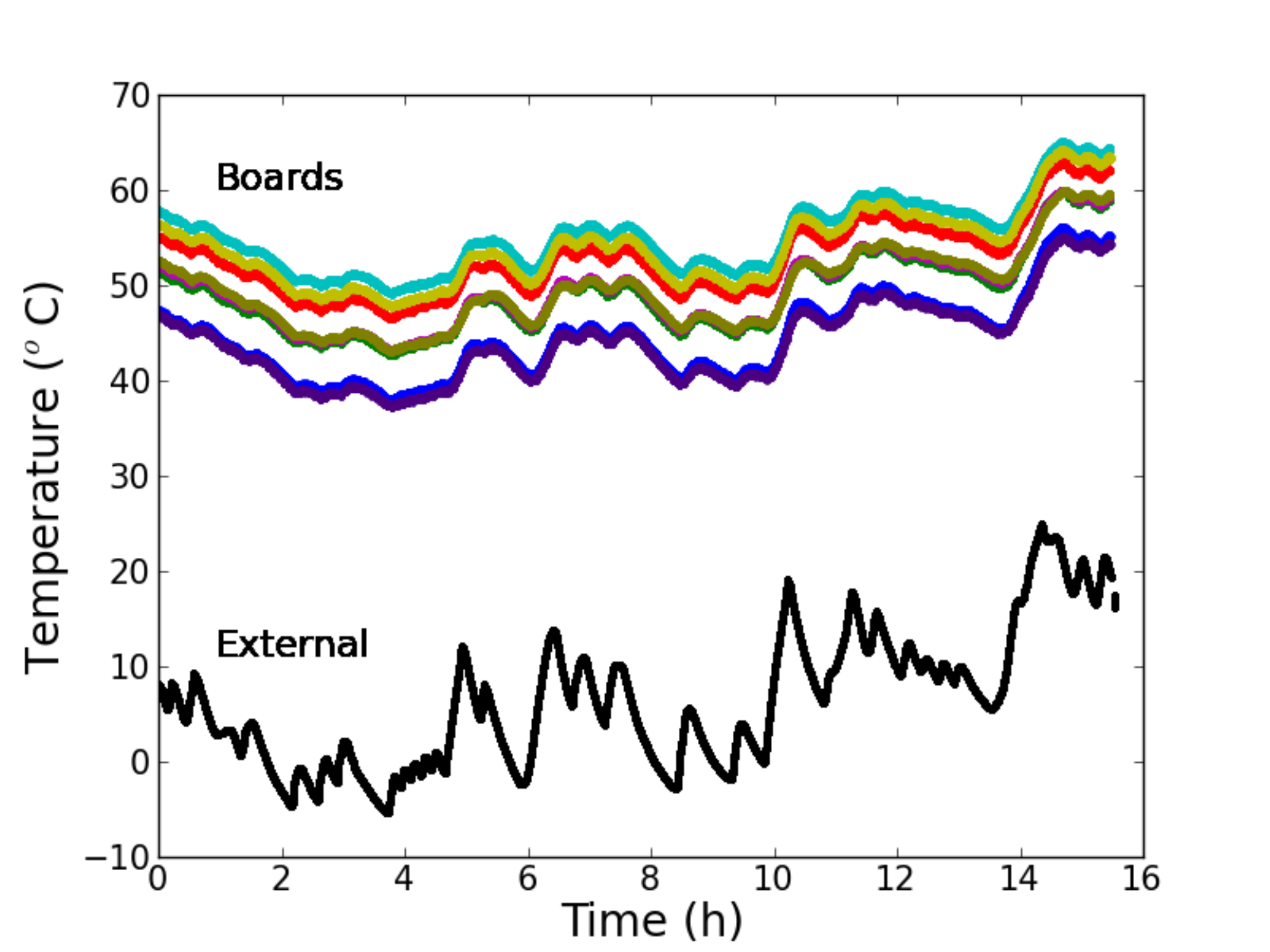}
\caption{Temperature of a \ac{BRC} top plate (``External'') and of the eight enclosed \ac{DfMUX} boards over a 16-hour period.  
Fluctuations of the External sensor strongly correlate with Sun orientation. The boards' temperature fluctuations are a low pass filtered 
version of the External sensor's due to the thermal conductance and heat capacitance of the liquid coolant. 
The warmer boards were located toward the middle of the crate, while the cooler boards were located closer to the ends of 
the crate.  Only five of the board temperature traces are clearly visible in this plot, as three traces overlap closely with others.} 
\label{figure: brc_temp}
\end{center}
\end{figure}

During flight the \ac{LCS} kept the readout boards within the required temperature range despite periodic exposure of the radiator panels 
to direct sunlight due to the gondola rotational motion (as described in Section \ref{sec:attitude_control}). Figure~\ref{figure: brc_temp} 
shows the temperature of the \ac{BRC} top plate along with the temperature of the enclosed \ac{DfMUX} boards during a representative 
segment of the flight. The nearly $10~^{\circ}$C observed difference between the warmest and coolest \ac{DfMUX} boards is due to the 
difference in radiative environment for the inner and outer boards inside the crate. The boards in an 8-board (6-board) \ac{BRC} were warmer 
by approximately $40~^{\circ}$C ($30~^{\circ}$C) relative to the exterior of the crate \citep{MacDermid_thesis}. 

\section{Attitude Determination and Control}
\label{sec:attitude_control}

\begin{figure}
\begin{center}
\begin{tabular}{c}
\includegraphics[width=0.6\columnwidth]{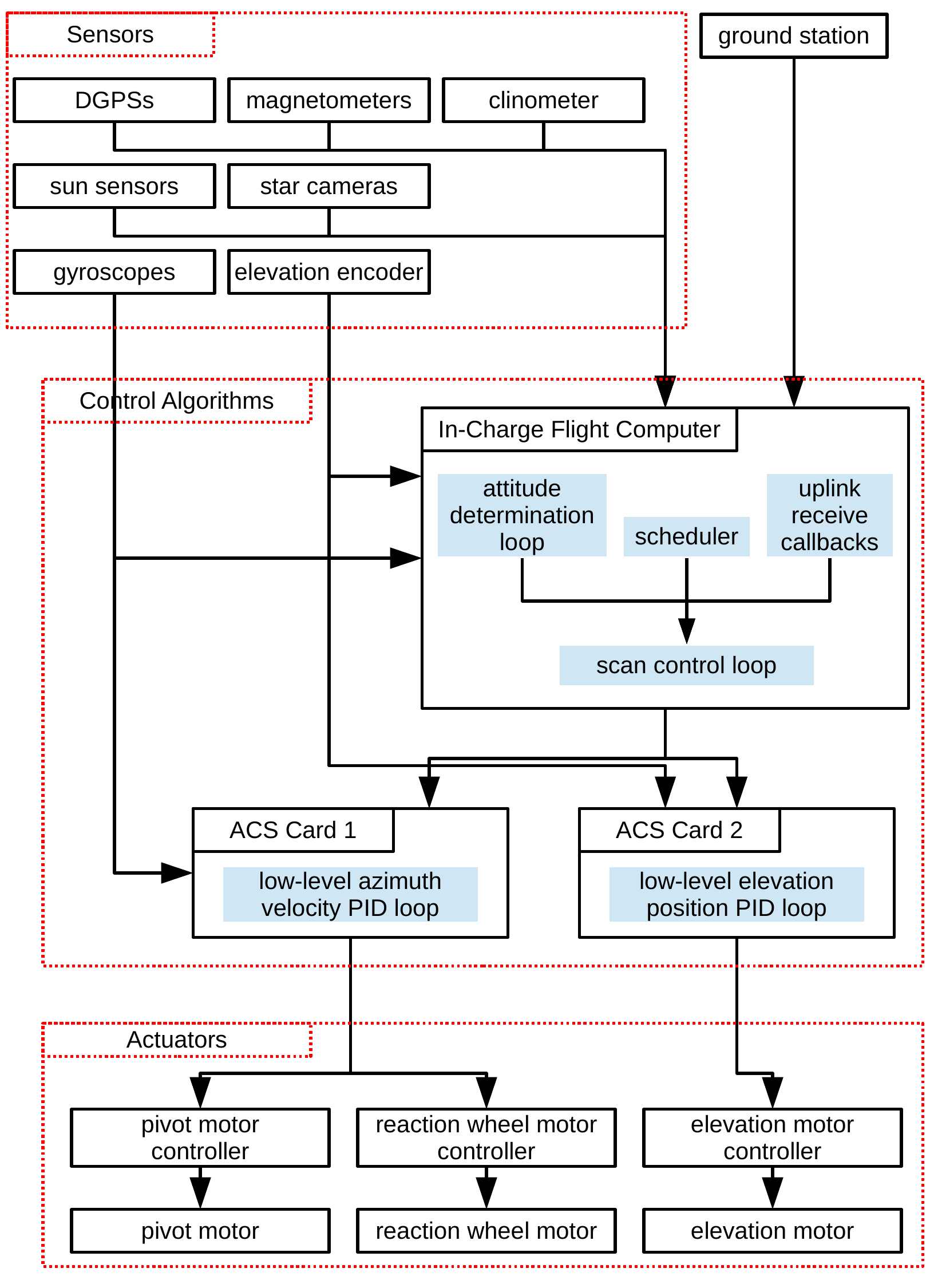}
\end{tabular}
\end{center}
\caption{
    The \ac{EBEX} \ac{ACS} consisted of three main components (red boxes): attitude sensors, control algorithms, and actuators. The sensors measured the 
instantaneous attitude, either absolute or relative to the gondola. The control algorithms processed this information to estimate the telescope attitude, 
compare it to a desired attitude as determined by a user-defined scan strategy, and issue instructions to actuators. Filled blue boxes represent 
software/firmware loops. Black bordered boxes represent physical components. Arrows represent flow of data.}
\label{figure: acs_diagram}
\end{figure}

The \ac{ACS} consisted of sensors, actuators, and a set of control algorithms operating in a feedback loop. Its role was to determine 
the instantaneous attitude of the telescope and execute a pre-defined sky scan pattern. It also acquired and stored the data required for post-flight 
attitude reconstruction. A block diagram of the \ac{ACS} is shown in Figure~\ref{figure: acs_diagram}. We present the main elements 
of the system, focusing on the sensors and the control algorithms.  We summarize the performance of the \ac{ACS} during the 2013 flight and the post-flight attitude reconstruction. 
The actuators are described in Section~\ref{sec:gondolamotion}. Elements of the \ac{EBEX} \ac{ACS} are  
also described in several additional publications~\citep{joy_ieee_paper,chappy_ieee_paper,joy_thesis,chappy_thesis}. 

\subsection{Attitude Sensors}

The \ac{ACS} sensors consisted of two redundant star cameras, two redundant 3-axis rate gyroscopes, 
two Sun sensors, two redundant magnetometers, a \ac{dGPS}, an elevation 
encoder, and an inclinometer. We also had access to information provided by the \ac{dGPS} of the \ac{CSBF}. 
Table~\ref{table: ebex_sensor_accuracy} summarizes the specifications and model of each sensor. 
\begin{table}[ht!]
\begin{centering}
\scriptsize
\begin{tabular}{llcrrr}
    \multicolumn{6}{c}{Attitude Sensor Specifications}     \\ \hline
    \hline
    Sensor & Model & Quantity   & Direction of      & In-flight   & Sample \\
           & name  & flown      & attitude provided & precision  & rate   \\
    \hline
    Star Camera & Kodak KAF-1603E       & 2  & az, el  & 1.3\as & Up to 0.5 Hz \\
                & Canon EF 200 mm f/1.8 &    & \& roll & 57\as  &  \\ 
    Magnetometers & Meda TFS 100 &  2  & az & 1$^\circ$ & 5 Hz \\
    Sun Sensors & Hamamatsu S5991-01 & 2  & az & 0.8$^\circ$ & 5 Hz \\
    Differential GPS  & Thales ADU 5 & 1  & az & 0.5$^\circ$ & 5 Hz \\
    Inclinometer & Geomechanics 904-T & 1  & el & 0.5$^\circ$ & 100 Hz \\
    Encoder  & Gurley A25S & 1  & el & 0.2$^\circ$ & 100 Hz \\
    Gyroscopes & KVH DSP 3000 & 6 & 3-axis rates & 40\as s$^{-1}$ & 1000 Hz \\
    \hline
\end{tabular}
\caption{List and specification of attitude sensors. The `precision' gives the standard deviations of the distributions in Figure~\ref{fig: star camera solutions} for the
star camera, and Figure~\ref{fig: other sensors} for other sensors. }
\label{table: ebex_sensor_accuracy}
\end{centering}
\end{table}

The primary sensors used for both real time control and post-flight attitude reconstruction were the star cameras and gyroscopes. The star cameras were mounted on 
either side of the inner frame and were approximately aligned with the telescope beam. The two gyroscope boxes, each consisting of three nearly orthogonal fiber optic gyroscopes, were also mounted on the inner frame. 

\begin{figure}
\begin{center}
\begin{tabular}{cc}
\makebox[\columnwidth][c]{
\centering \includegraphics[width=0.5\columnwidth]{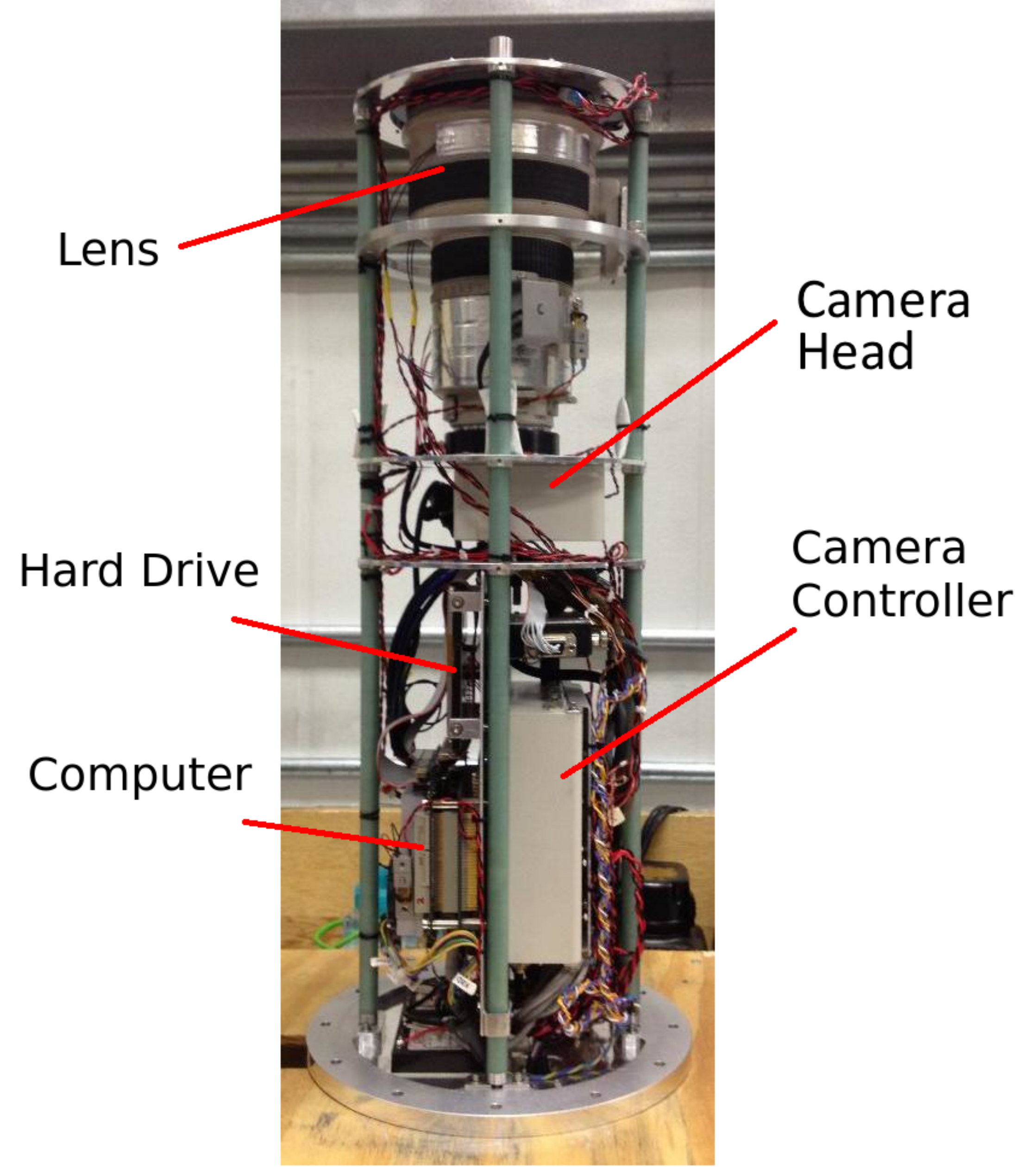} 
\centering \includegraphics[width=0.5\columnwidth]{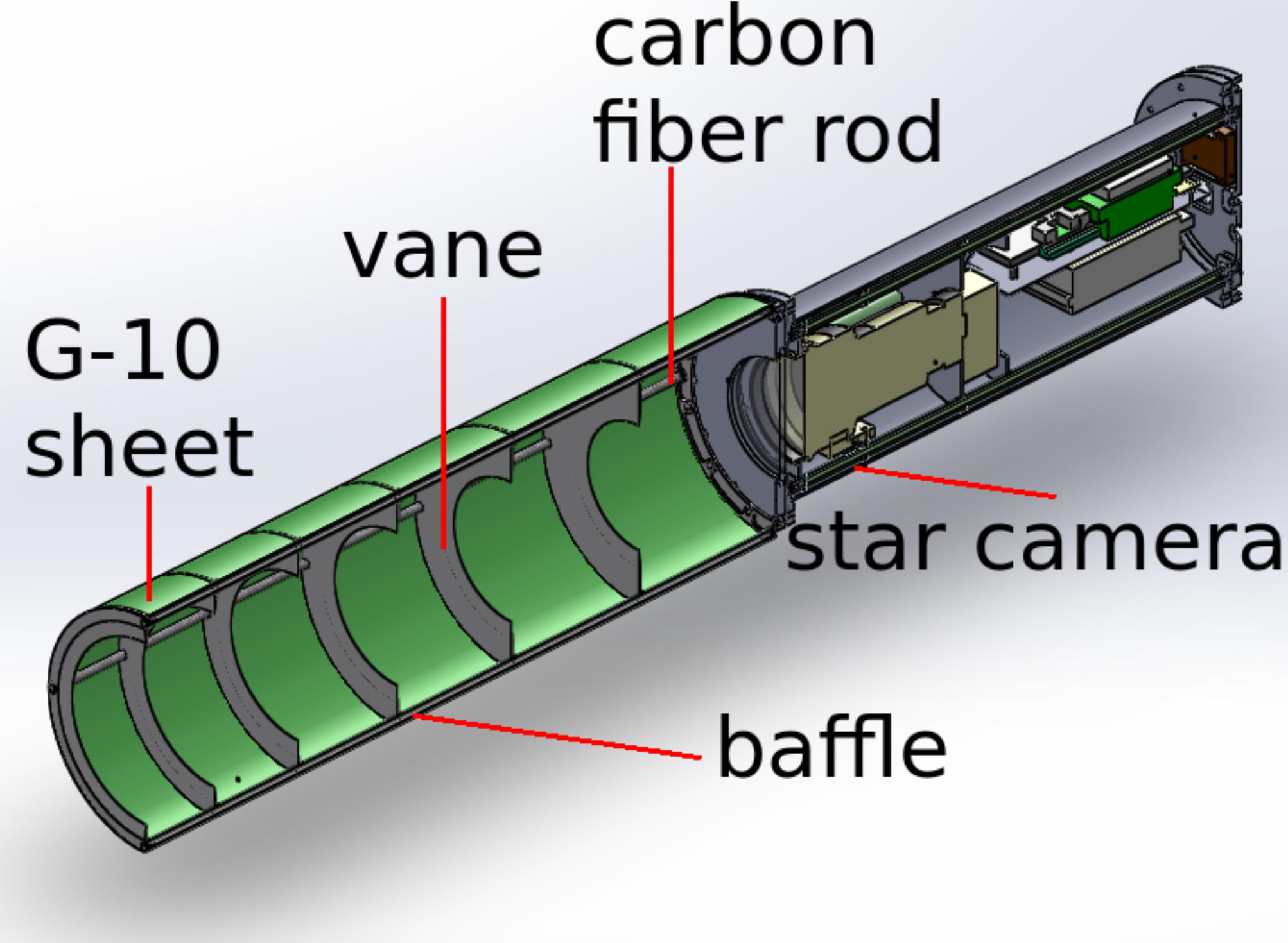}}
\end{tabular}
\end{center}
\caption{The star camera assembly consisted of a pressurized vessel that held the star camera hardware and a baffle (right). 
The baffle was made of a 87.6~cm long tube of G-10 fiberglass sheet that was wrapped around thin aluminum vanes connected with carbon fiber tubes. The baffle weighed 1.87~kg. Inside the vessel (left), which was 
pressurized with N$_{2}$ gas to 1~atm, were the camera head, lens, camera controller, and computer. } 
\label{fig: xsc}
\end{figure}

Each star camera consisted of a telephoto lens, a filter,\footnote{Red color 25A filter from Hoya Filters} 
a \ac{CCD} camera, and a computer mounted in a rigid assembly inside a cylindrical vessel filled with nitrogen 
gas at a pressure of 1~atmosphere.  A stray light baffle was mounted to the exterior of the star camera pressure vessel; see Figure~\ref{fig: xsc}. All parts inside the baffle were painted with flat black spray paint.\footnote{Krylon Ultra-Flat Black} 
The star camera found an attitude by taking a picture of a star field and comparing the image to a star catalog. 
The two star cameras were redundant to ensure that attitude solutions were available even if one failed.  
The star camera computer ran the \ac{STARS}, a platform-independent software 
custom-written for \ac{EBEX} in C++ that captured the images, found the bright spots in the image, matched their pattern to a known 
catalog of stars, and communicated the resulting solution to the flight control program (FCP) operated by the main flight computer~\citep{chappy_spie}. 
\ac{STARS} was optimized to find stars even when the camera was out of focus or when the stars were blurred due to gondola motion. 
Each camera had a point spread function with a full width half maximum of 9\as and a field of view of $4.05^\circ~\times~2.70^\circ$. 
Star camera exposure time was set to 300~ms in order to reliably detect stars 
with apparent magnitude 7.3 or brighter. With this exposure time, the motion blur solving feature of \ac{STARS} permitted the cameras 
to solve images taken with azimuthal velocities up to $0.02^\circ$~s$^{-1}$. 

The star cameras performed well during flight, consistently solving the images in real-time with minimal intervention. 
The \ac{STARS} software overcame several unanticipated challenges: \\ 
(a)  The loss of azimuth control (see Section~\ref{section: acs_flight_and_control}) prevented \ac{STARS} from performing the autofocus algorithm, which 
required stationary pointing, and both star cameras were slightly out of focus during the entire flight. 
\ac{STARS} continued to find stars in the images, however, because of its robust source detection algorithm. \\
(b) To solve images quickly, \ac{STARS} normally used a coarse attitude determination by the secondary sensors. 
The coarse approximation minimized the search radius when matching the stars in the image to the catalog of stars. 
The \ac{dGPS} failed to provide information for multiple sections of the flight, which prevented the attitude guess from 
the secondary sensors to be transformed from the local az/el reference frame to the equatorial reference frame in which the 
cameras operated. Yet even in those sections \ac{STARS} continued finding solutions within several seconds, switching to its `lost-in-space' mode. 
The \ac{STARS} catalog was optimized by pre-computing the distances between combinations of stars and by filtering the catalog down to  
fewer than 20 stars per field of view. 
Without the optimizations implemented in \ac{STARS}, finding solutions without directional guidance could take a few minutes per image~\citep{chappy_thesis}. \\   
(c) The \ac{STARS} software successfully identified stars and matched stellar patterns in the presence of image non-idealities, including 
passing polar mesospheric clouds, vignetting, and internal reflections~\citep{chappy_thesis}.

\begin{figure}[ht!]
    \centering
    \includegraphics[width=0.7\columnwidth]{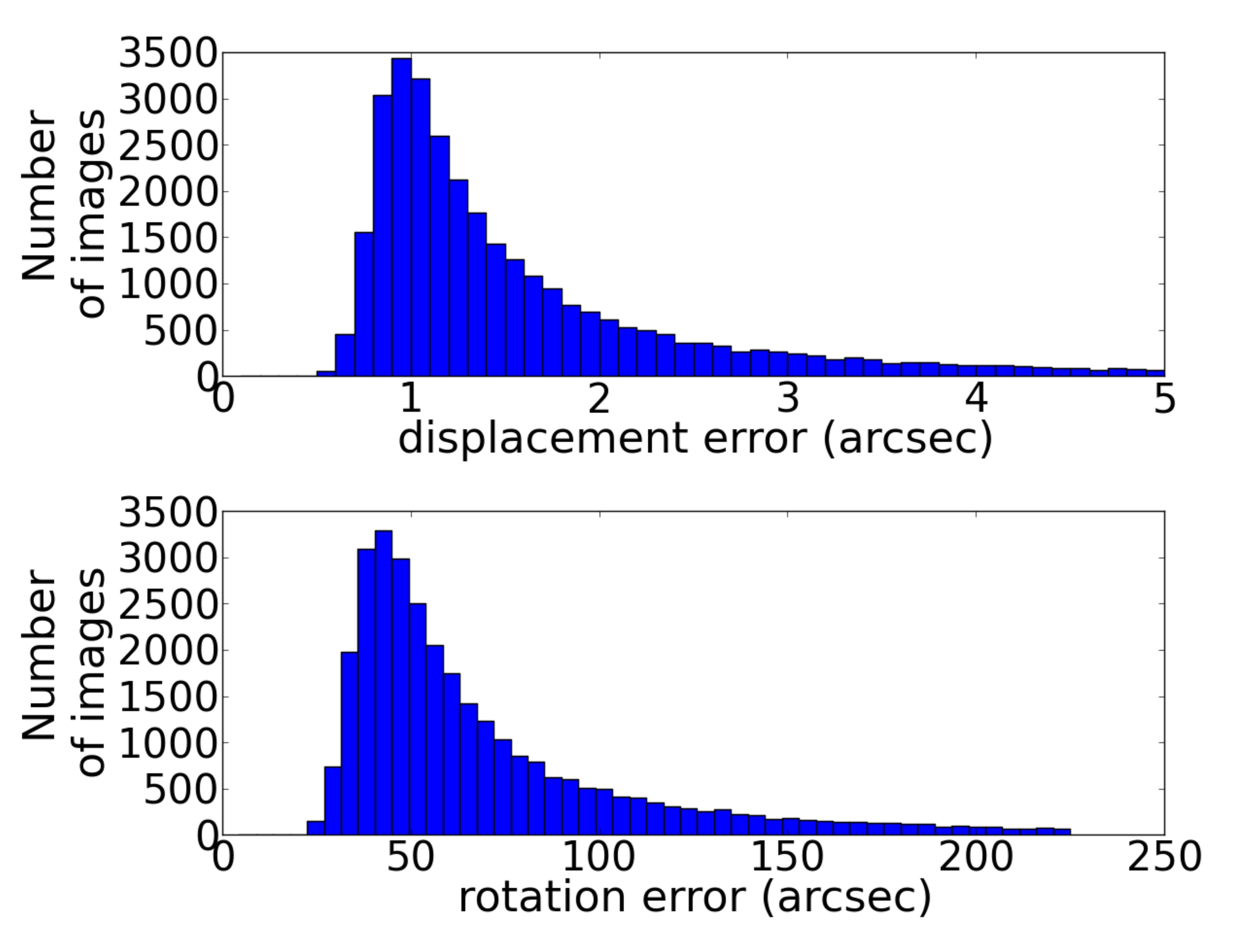}
    \caption{Histogram of the attitude solution uncertainty as reported by the pattern matching least square algorithm for all solved images. 
    The top panel shows the displacement uncertainty, i.e., the combined uncertainty from RA$\times$cos(Declination) and Declination. 
    The bottom panel shows the rotation uncertainty around the image center. The median uncertainty is 1.3\as in displacement and 57\as in rotation. }
    \label{fig: star camera solutions}
\end{figure}

The two star cameras acquired a total of 41,262 images, 80\% of which provided attitude solutions post-flight. Most of the remaining images were saturated 
because the cameras were pointing at the balloon during housekeeping operations or the attitude was within 30$^\circ$ of the Sun. 
On average there were 8 stars per image. 
Figure~\ref{fig: star camera solutions} shows a histogram of the solution uncertainty from all solved flight images. The uncertainty was reported 
by the least square algorithm matching the pattern of stars in the image to the catalog.  

Each gyroscope measured the rate of angular rotation around its axis, and outputted a digital signal at 1000 Hz which was read out by 
an on-board \ac{DSP} unit.\footnote{Provided by the University of Toronto} The data were despiked, passed through a box-car infinite impulse response filter with a cutoff frequency of 20~Hz, and written 
to disk at 100~Hz. The gyroscopes were chosen for their combination of cost, relatively low white noise (40\as~s$^{-1}$), and their bias timescale of $\sim$200~s. 
Three gyroscopes were mounted inside a precision machined aluminum box with connecting surfaces orthogonal to within 5\am.
The gyroscopes were wrapped in overlapping strips of magnetic shielding\footnote{Metglas, Inc.} to reduce their susceptibility to ambient magnetic fields. 
The shielding reduced the gyroscopes' zero-motion bias from 17\as~s$^{-1}$~G$^{-1}$ to 3\as~s$^{-1}$~G$^{-1}$ \citep{Britt_thesis}. 
The gyroscopes performed well during flight, recording data continuously and exhibiting 
white noise and bias behavior in accordance with pre-flight measurements.

\begin{figure}[ht!]
\centering \includegraphics[width=0.8\columnwidth]{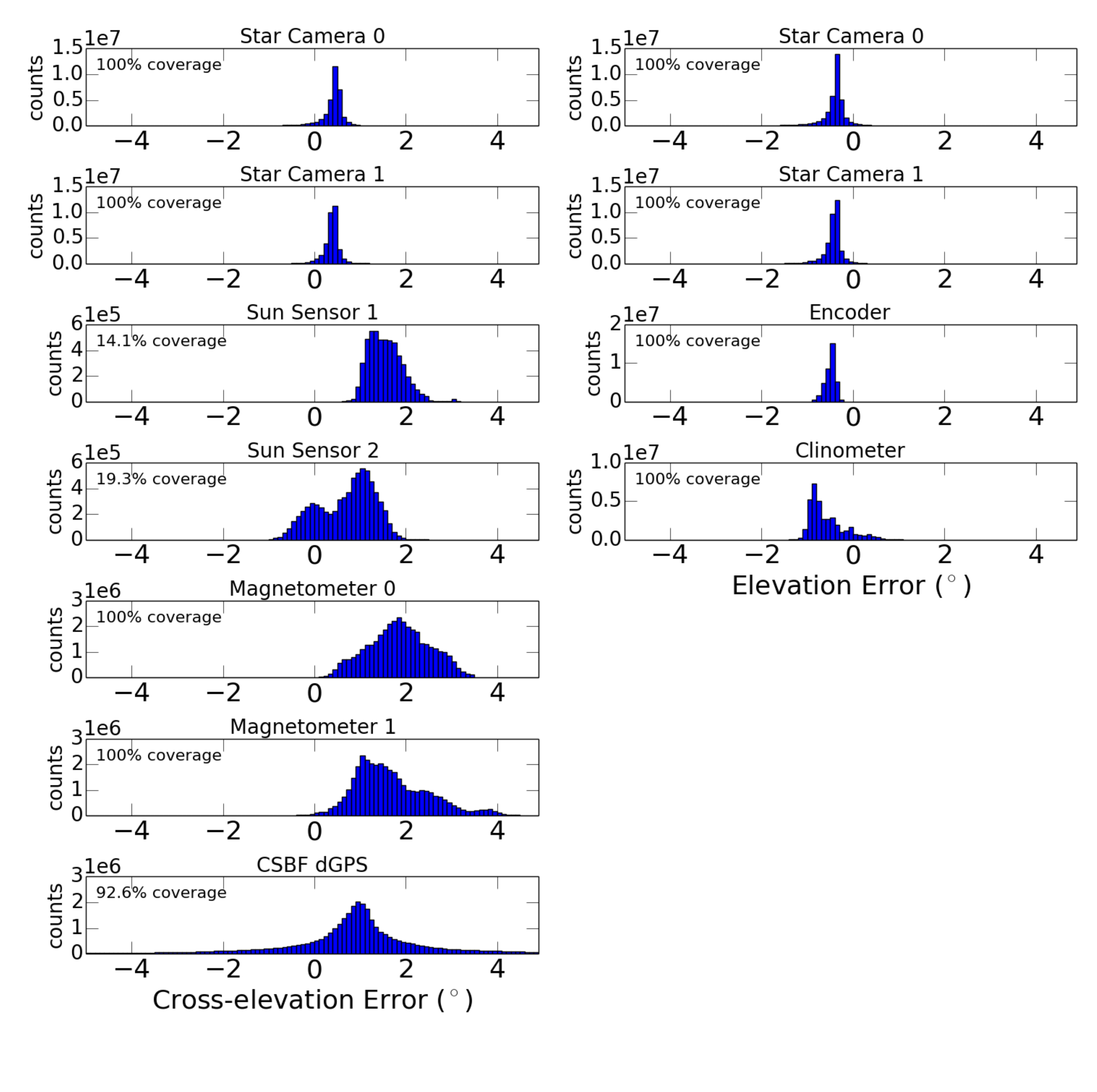} 
\caption{The performance in cross-elevation (defined as azimuth$\times$cos(elevation), left column) 
and elevation (right column) of each absolute pointing sensor during the EBEX2013 flight given the {\textit{pre-flight}} calibration procedure. Cross-elevation is calculated using the post-flight elevation.
Each plot shows a histogram of the difference between the post-flight reconstructed boresight attitude, and the in-flight sensor 
attitude computed using pre-flight offset calibration. Coverage values give the percentage of time the sensor provided valid attitude.
The \ac{EBEX} \ac{dGPS} is not plotted as it failed early in flight and did not provide attitude.} 
\label{fig: other sensors}
\end{figure}

As listed in Table~\ref{table: ebex_sensor_accuracy}, \ac{EBEX} also flew a complement of secondary sensors designed to provide coarse 
real-time attitude to be used as a pointing guess for the star cameras, and intended to provide back-up in case the star cameras failed to 
solve images real-time. The main source of error for real-time attitude determination using the coarse sensors was the calibration of each 
sensor's overall directional offset. Before flight, we measured these offsets by referencing the 
sensors to star camera solutions obtained using the few stars bright enough to be visible by the star camera from the ground during the Antarctic summer.  
Directional offsets were re-calibrated periodically in-flight using the star camera solutions. Figure~\ref{fig: other sensors} gives the in-flight 
performance for each of the secondary sensors given the calibration performed {\textit{pre-flight}}  
and ignoring all in-flight re-calibrations. This is a `worst-case-scenario' indicating what the performance of the sensors would have been 
had the star cameras not provided any re-calibration during flight. The dispersion 
about the mean of each sensor is an indicator of each sensor precision over more than 10 days of flight, and the mean of each sensor is 
an indicator of the accuracy of the {\textit{pre-flight}} calibration.

\subsection{Control Algorithms}
\label{acs: control loops}

Three sub-programs operated in a feedback loop to control the instrument attitude (see Figure~\ref{figure: acs_diagram}): the attitude determination
sub-program used sensor information to estimate the telescope attitude; the scan pattern sub-program determined the instantaneous desired attitude 
and scan rate; and the low-level sub-program sent current to the azimuth and elevation actuators. The attitude determination and scan pattern sub-programs 
ran on the `in-charge' flight computer -- one of the two redundant flight computers (see Section~\ref{sec:computers}) -- at 100.16~Hz. The azimuth and 
elevation low-level sub-programs ran on two \ac{DSP} cards\footnote{ADSP 21062 from SHARC by Analog Devices, Inc.} at 10,400~Hz. 

The attitude determination sub-program estimated the telescope attitude by performing a weighted average of the information 
obtained from all sensors deemed 
operational by ground operators. Horizontal roll was approximated as zero. Each sensor's attitude information was estimated using a 
1-D Kalman filter that evolved the sensor prior attitude using the gyroscopes' data and included new available measurements.

The primary scan pattern was a raster scan. The algorithm to perform this scan was a state machine that alternated between 
scanning at constant azimuth velocity, pausing to capture star camera images, and stepping to the next elevation. Given the scan parameters 
and the current attitude, the algorithm output was a target azimuth velocity and target elevation position at every time step. 

The requested velocities and attitudes were transmitted to the \ac{DSP}s which had proportional-integral (PI) feedback loops operating 
on the difference between current and target quantities \citep{joy_thesis}. The outputs of the PI loops were 
ultimately converted to a \ac{PWM} signal for the motor controllers. The PI values were tuned in-flight to ensure optimal motion 
of the telescope. The feedback loops had override modes that allowed the ground operators to command \ac{PWM}s manually. In flight, we 
employed both manual \ac{PWM}s and the automatic scan algorithms.

\subsection{In-Flight Performance}
\label{section: acs_flight_and_control}

\begin{figure}[ht!]
\begin{tabular}{cc}
\includegraphics[width=0.5\columnwidth]{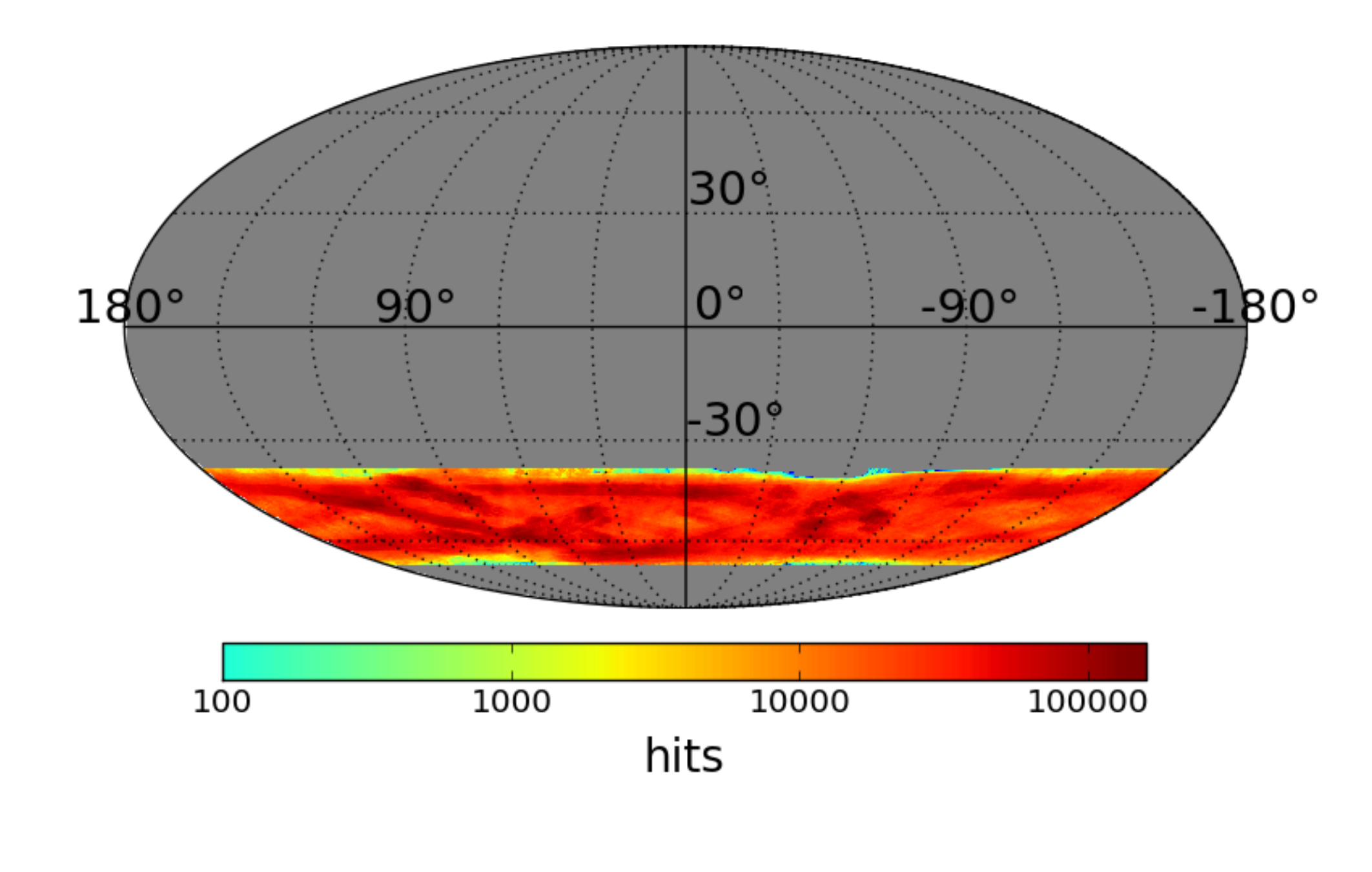}
\includegraphics[width=0.5\columnwidth]{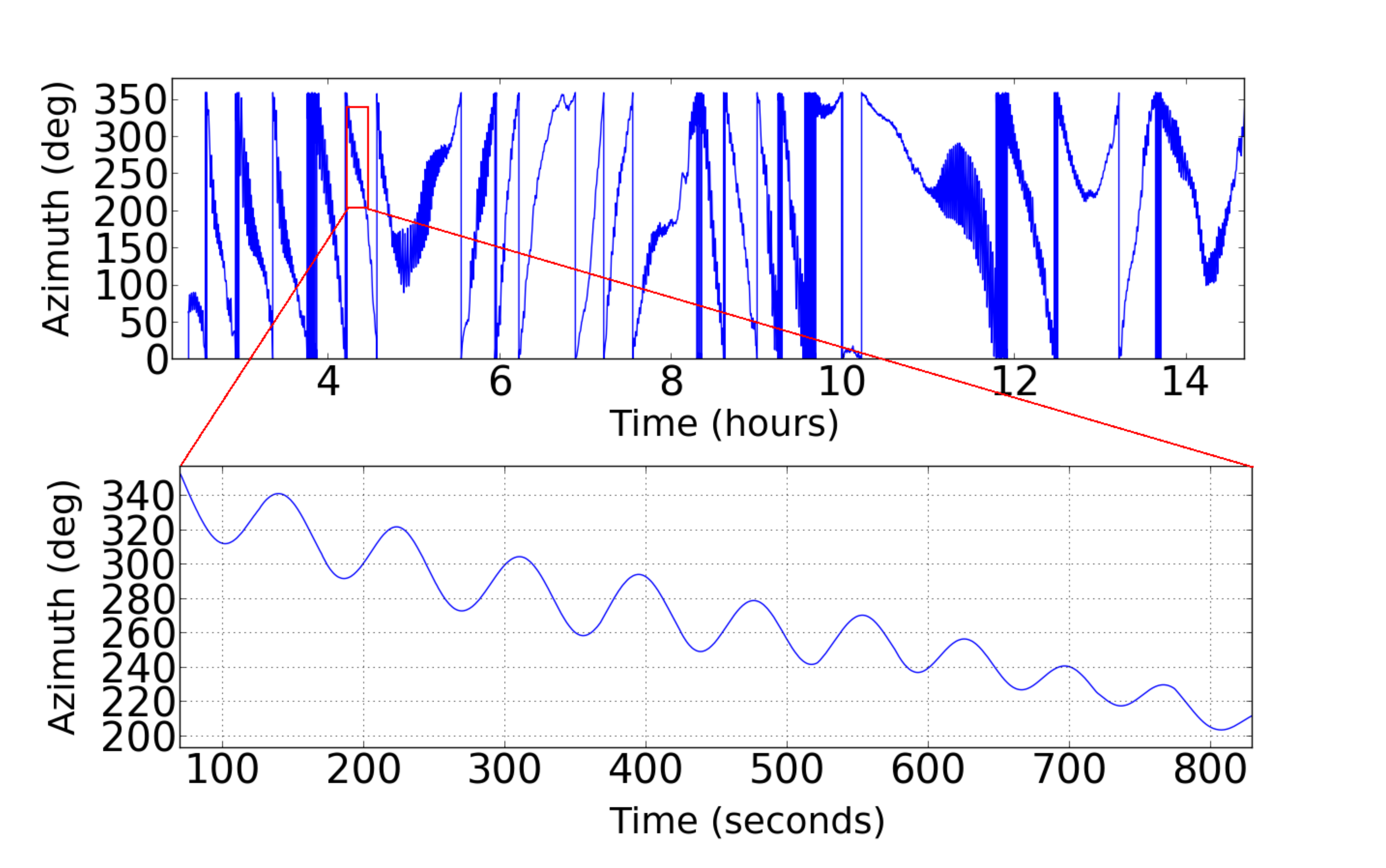}
\end{tabular}
\caption{{\it Left}: A map in equatorial coordinates of the number of detector samples per pixel (hit map) for the EBEX2013 flight from all frequency bands. {\it Right:} Typical patterns in the azimuth motion during the EBEX2013 flight. Over long time scales the gondola executed 
full $360^\circ$ rotations with occasional reversal of direction (top panel). Superposed was 
an oscillatory motion (bottom panel) with 80~s period. This period matched the one predicted given the moment of inertia of the gondola and the 
torsional constant of the flight line. This rotational motion had variable amplitude that reached up to tens of degrees. }
\label{fig: az motion and hitmap}
\end{figure}

The \ac{EBEX} payload launched from McMurdo, Antarctica on December 29, 2012. It circumnavigated the continent, taking data for 11 days at 
an average altitude of 35~km. Shortly after reaching float altitude we discovered that the pivot motor controller was overheating and shutting down 
(see Section~\ref{sec:thermaloverview}). Without active control, the azimuth of the gondola was determined by the rotation of the balloon 
and the rotational spring constant of the flight train. The resulting azimuth motion is shown in Figure~\ref{fig: az motion and hitmap}. 
It was a superposition of 
full rotations with variable rotational speed and 80~s period oscillations that had variable amplitude.  Throughout 
the flight, more than 97\% of the azimuthal speeds were below 1$^\circ$~s$^{-1}$. 
We oriented the gondola at constant elevation of 54$^\circ$ in order to maintain an angular 
separation of $\sim$15$^\circ$ between the telescope boresight and both the balloon and the Sun's maximum elevation. 
The resulting sky 
coverage was a strip of sky delimited by declination $-67.9^\circ$ and $-38.9^\circ$, covering an area of 5700~square degrees; 
see Figure~\ref{fig: az motion and hitmap}. By a 
fortunate coincidence, the 80~s natural rotational oscillation period of the gondola and flight-line matched the designed scan strategy. Thus 
the gondola came to a stop every $\sim$40~s, enabling star camera images to be taken while the gondola was in the 
stationary position that is optimal for star camera imaging. 
In this manner, all the pre-flight work of assessing attitude determination accuracy was still relevant to the actual scan pattern of the EBEX2013 flight. 

\subsection{Post-Flight Attitude Determination}
\label{sec:post-flight attitude determination}

Errors in attitude determination convert {\it E}-mode to {\it B}-mode signal. To keep these spurious {\it B}-modes negligible, we placed a 
requirement that the spurious {\it B}-modes would be less than 10\% of an inflationary {\it B}-mode with $r=0.05$ and nominal cosmology lensing signal
within $ 30 \lesssim \ell  \lesssim 1500$, which was the range the instrument was designed to probe. 
\citet{hu03} quantified the effects of several types of experimental errors, including attitude errors, on the 
determination of the {\it B}-mode power spectrum. In their formalism, attitude errors are characterized in terms of their spatial power spectral density, 
and the induced {\it B}-modes are given in terms of a convolution with the cosmological {\it E}-modes. 
In this Section we discuss the approach we used to reconstruct attitude post-flight and 
quantify the attitude errors. We refer to the entire pipeline as \ac{ADS}. \citet{joy_thesis} used the \ac{ADS} to construct the
spatial spectral density of the measured attitude errors, convolved it with the cosmological {\it E}-modes, and showed that the requirement 
on attitude reconstruction for the EBEX2013 flight has been met.  

Attitude errors grow with time between star camera readings because of gyroscope rate noise and uncertainties in the \ac{TM} between 
star cameras and gyroscopes. Although using the  combination of star cameras and gyroscopes is common on pointed balloon-borne 
instruments, the employment of an extended 40~s scan between star camera images necessitated a detailed analysis of the system 
through simulations and the development of a judicious post-flight \ac{ADS} to ensure that attitude errors met the requirement.  

Star camera images provided attitude solutions that far exceeded the requirement. Between times for which images were available 
we integrated data from the gyroscopes; we refer to this as the \ac{IA}. Attitude errors for the \ac{IA} originated from gyroscope 
slow-varying noise, which were a function of time, and from a time-independent inaccuracy in the \ac{TM} between the gyroscopes 
and the star cameras' frames of reference. There were two contributors to inaccuracy in the \ac{TM}: an inaccuracy in the alignment 
matrix that orthogonalized the gyroscopes -- their hardware mounting was not perfectly orthogonal -- and inaccuracy in the rotation 
matrix that rotated this orthogonalized frame to align with the star camera frame. Priors on the \ac{TM} were obtained using pre-flight 
measurements of the gyroscope box orthogonality.

\begin{figure}[ht!]
\centering
\includegraphics[width=0.9\textwidth]{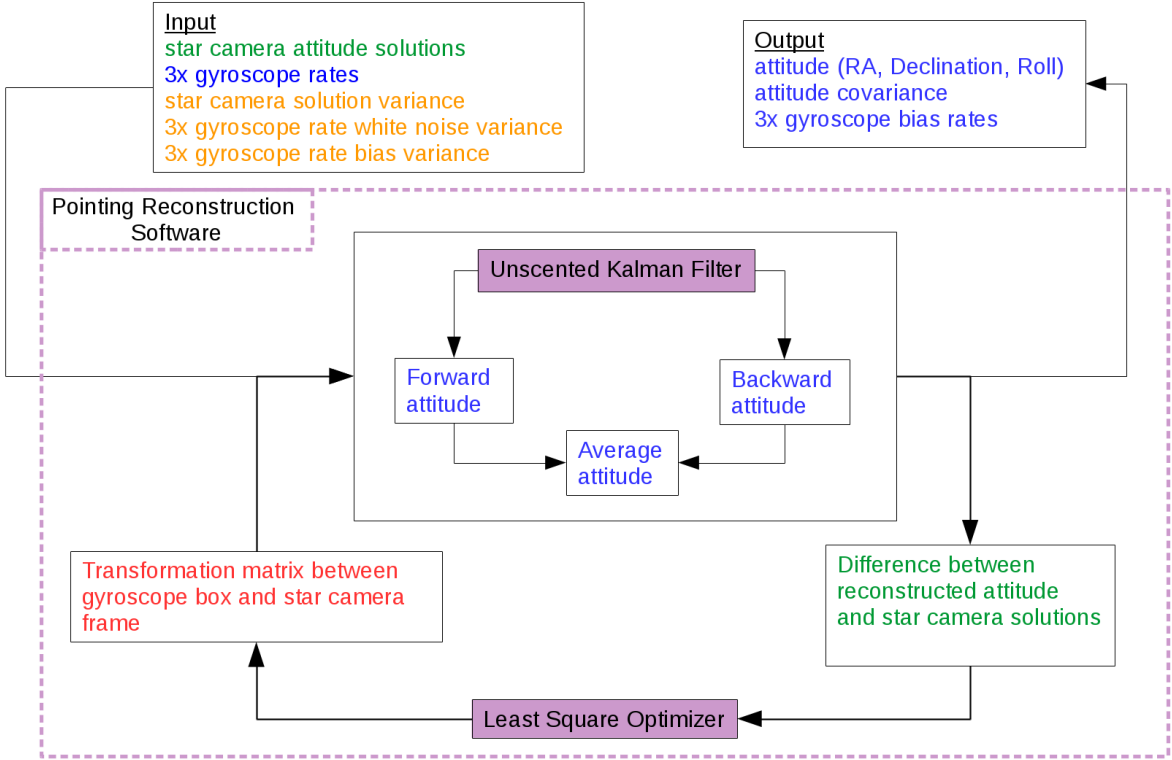}
\caption{Diagram of the \ac{ADS}. The filled purple boxes denote the two primary constituent codes. The blue text designates 
arrays of data with identical lengths and a sample rate of 100.16~Hz. The green text represents much smaller arrays with 
length equal to the number of star camera solutions. The yellow text designates single numbers, and the red box is a $3\times3$ matrix with six independent parameters.}
\label{figure: pointing overview}
\end{figure}

The ADS found both time-dependent and time-independent parameters through an algorithm that combined an 
Unscented Kalman Filter (UKF)~\citep{ukf} and a least square optimizer, as shown in Figure~\ref{figure: pointing overview}. 
Using a given \ac{TM}, the UKF determined the attitude and estimated the slowly time-varying gyroscope offsets. 
It ran forward and backward in time producing a forward and a backward \ac{IA}, as well as a solution that was the weighted 
average of the \ac{IA} in each temporal direction.
When each star camera measurement was made, the UKF computed the differences between the image solution and the forward and backward \ac{IA}. 
The least square optimizer iteratively minimized these differences over the entire 11~day flight to find the optimal parameters of the 
time-independent \ac{TM}. Each of the (multi-processed) 90 iterations required to reach convergence took the equivalent of 80~min on a single 2.1 GHz processor.
The error on the \ac{TM} rotation and misalignment angles, evaluated by simulating sensors' performance and reconstructing a 
known attitude and TM, were found to be within 3.4\am. 

\begin{figure}[ht!]
\centering
\includegraphics[width=0.5\columnwidth]{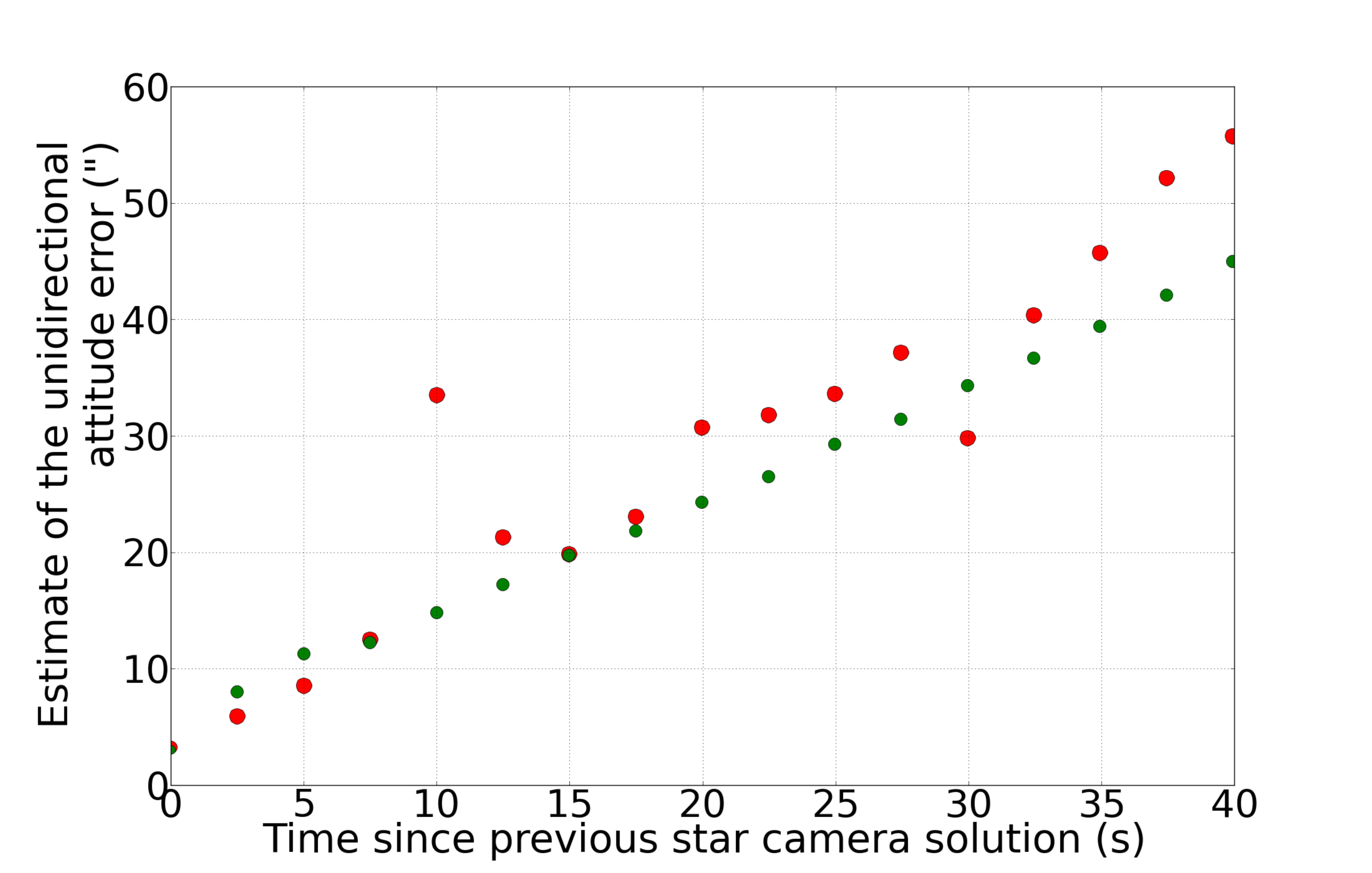}
\caption{Estimate of the unidirectional attitude error $\sigma_{\Delta t}$ as a function of the time $\Delta t$ since the last star camera solution. 
All throws are binned in 2.5~s bins. Data are shown up to $\Delta t = 40$ s because these encompass the majority of times. 
Values for the red dots are computed by collecting the differences between star camera solutions and unidirectional \ac{IA} for all star camera 
images that fall within that bin. The value plotted is the standard deviation of the distribution in that bin. In green is plotted the 
unidirectional error estimated by the UKF, showing agreement with the measured data points.}
\label{fig: flight_reconstruction}
\end{figure}

We evaluated the increase in attitude errors as a function of time separation $\Delta t$ since the last star camera solution in the following way. 
For the forward or backward \ac{IA}, which we call unidirectional \ac{IA}, the error grew until a new star camera image was included in the solution, 
and the error at that sample was estimated using the difference between the unidirectional \ac{IA} and the star camera solution, 
before the latter was included in the \ac{IA}. 
We measured the error as a function of $\Delta t$ by using pairs of star camera readings separated by that time. 
For each $\Delta t$, in bins 2.5 s wide, we histogrammed the differences between the star camera solution and that given by the \ac{IA}. 
We included both forward and backward \ac{IA} data points. The distribution means were near zero, but the standard deviations of the 
distributions  $\sigma_{\Delta t}$ gave an estimate of the unidirectional attitude error at $\Delta t$ away from a star camera solution. 
Figure~\ref{fig: flight_reconstruction} shows the unidirectional error $\sigma_{\Delta t}$ as a function of $\Delta t$ for the EBEX2013 flight.

For the average attitude solution -- constructed from the forward and backward \ac{IA} -- the 
attitude error $\sigma^A_{\Delta t} $ at any time since the last star camera solution 
was calculated using the unidirectional errors via:
\begin{eqnarray}
\label{eq: xsc rms}
\frac{1}{(\sigma_{\Delta t}^A)^2} = \frac{1}{\sigma_{\Delta t}^2} + \frac{1}{\sigma_{(T-\Delta t)}^2} 
\end{eqnarray}

\noindent where T was the total time between star camera images.  The error was largest mid-throw and decreased close to the times the images were taken. 
\citet{joy_thesis} shows that, when the attitude errors of the average solution 
are translated into the $\ell$ domain using the EBEX2013 scan strategy, the spurious {\it B}-mode generated is less than 1/10 of the 
CMB lensing power spectrum for $\ell \leq 1500$. 

\section{Flight Management}
\label{sec:flightmanagement}

The EBEX2013 flight marked the first use of a kilo-pixel array of \ac{TES}
bolometers aboard a balloon-borne payload. The short observation 
time available, the limited telemetry and commanding bandwidth, the modest available power and computing resources relative to a 
ground-based experiment, the large throughput of data, and the complexity of operating a kilo-pixel array required 
the development of (1) an efficient method to tune and control 
the \ac{TES} bolometers; (2) specialized software to collect, store, and telemeter data; and (3) an on-board scheduling system
to manage the multitudes of automated tasks that had to take place.  Here we describe
how we solved these challenges. More details are provided by \citet{MacDermid_thesis} and \citet{Hillbrand_thesis}. 

\subsection{On-Board Computers} 
\label{sec:computers}

We used two redundant, ruggedized, low power single board computers.\footnote{AMPRO computers by ADLINK Technology, Inc.}  Each computer
had a 1.0~GHz Celeron processor, 256~MB RAM, and a 1~GB solid state disk. The computers operated in 
ambient pressure and had a steady state power consumption of 19.5~W each. 
The solid state disk stored the computer operating system, additional modular drivers, and the 
\ac{FCP}, which was configured to run immediately after
the computer booted. The \ac{FCP}, originally inherited from BLAST \citep{wiebe_thesis} and heavily modified, controlled all aspects of payload operation including scheduling observations, collecting data from the bolometer readout boards and from various housekeeping systems, 
storing data on board and telemetering to the ground, receiving commands from ground-operators and distributing them to 
on-board subsystems, and occasionally triggering a set of commands that were pre-programmed before flight. 

We developed an `event scheduler' that controlled all flight events. In its default mode, it controlled all on-board 
operations without operator interference. Various experiment events, such as planned sky observations and cycling of the 
sub-Kelvin refrigerators, were pre-programmed and referenced to \ac{UTC}. The detailed sequence of commands necessary to 
conduct, for example, sky observations or a refrigerator cycle were stored in `schedule files' containing hundreds
of individual commands. When triggering an event, the event scheduler
launched and tracked the operation of the appropriate schedule file; when necessary, such as with sky observations, it operated with sidereal time. 
Ground operators also over-rode the default scheduling, uploading alternate schedule files or triggering various pre-determined schedule files. 

Flight computer redundancy was implemented via a watchdog card\footnote{Provided by the University of Toronto} connected to the IEEE-1284 parallel port of each computer \citep{wiebe_thesis}. 
In nominal operation the \ac{FCP} watchdog thread toggled a pin on the parallel port at 25~Hz. 
If this action ceased for more than 1~s a fault was inferred and the watchdog card power-cycled 
the faulty computer and switched control to the other computer.  
The identity of the computer in control was communicated to both flight computers via a common bus and recorded. 
During the 11-day flight we logged 19~non-commanded changes of the in-control computer, which we attribute to single event upsets. 
Aside from these occasional reboots, both computers operated throughout the flight. 

\subsection{Timing System} 
\label{sec:timing}

Data collected by the various flight subsystems, including flight computers, the attitude control system, the receiver housekeeping system, and the
detector readout system, were stamped by each subsystem asynchronously. 
We synchronized these subsystems using a common time system, called EBEXTime, described below.  Additional details are provided in \citet{Sagiv_thesis}.

The time synchronization system consisted of a time server and various time clients. Communication with the time server was handled via a \ac{CAN bus} card. 
There were two time servers on board for redundancy, thus there were EBEXTime1 and EBEXTime2.  
The boards were connected to the \ac{BRC}s and the \ac{ACS} electronics 
clients via an RS-485 serial line, and to the receiver housekeeping electronics and the flight computer clients via \ac{CAN bus}. 

EBEXTime is the number of 10~$\mu$s ticks since the start of a `major period', which was at most 6~hours duration. 
The major period counter was stored in non-volatile EEPROM on the timing server board.
A new major period started  
each time the time server was powered on. The flight computer could optionally set the time-server's major period upon power-up initialization. 
The full EBEXTime datum was a 48-bit word consisting of a 2-bit Board ID, a 14-bit major period register and a 32-bit tick counter. 
Each time client maintained its own copy of the EBEXTime.
Each client's 32-bit tick counter was incremented by a local oscillator at 100~kHz. 
The time server, at a rate of 6.1~Hz (0.16384~s = 214~ticks), broadcasted a synchronization message consisting of its board ID, 
the high 32~bits (i.e. the major period and the high 18~bits of the tick counter) to all clients. 
On receipt of a valid synchronization message the client rewrote its 46-bit time word with the 32 bits received plus 14 zero bits appended.
Local client clocks used oscillators with $\pm$25~ppm stability. The master clock on the time server used an oven-controlled oscillator with 
a temperature stability of $\pm$0.2~ppb between $-20~^{\circ}$C to $+70~^{\circ}$C. 

Upon power-up, clients used either of the time-servers' synchronization messages available on their bus. 
They automatically switched to the other if one became unavailable.   
Both time servers were synchronized to absolute time post-flight using the in-control flight computer CPU time. 
Because the entire attitude solution was conducted in equatorial coordinates and all data was co-stamped with EBEXTime together 
with the attitude solutions, only very coarse (tens of minutes) synchronization with absolute time was necessary. 

\subsection{On-Board Network} 
\label{sec:network} 

\begin{figure}[hpb!]
\begin{center}
\includegraphics[height=4.5in]{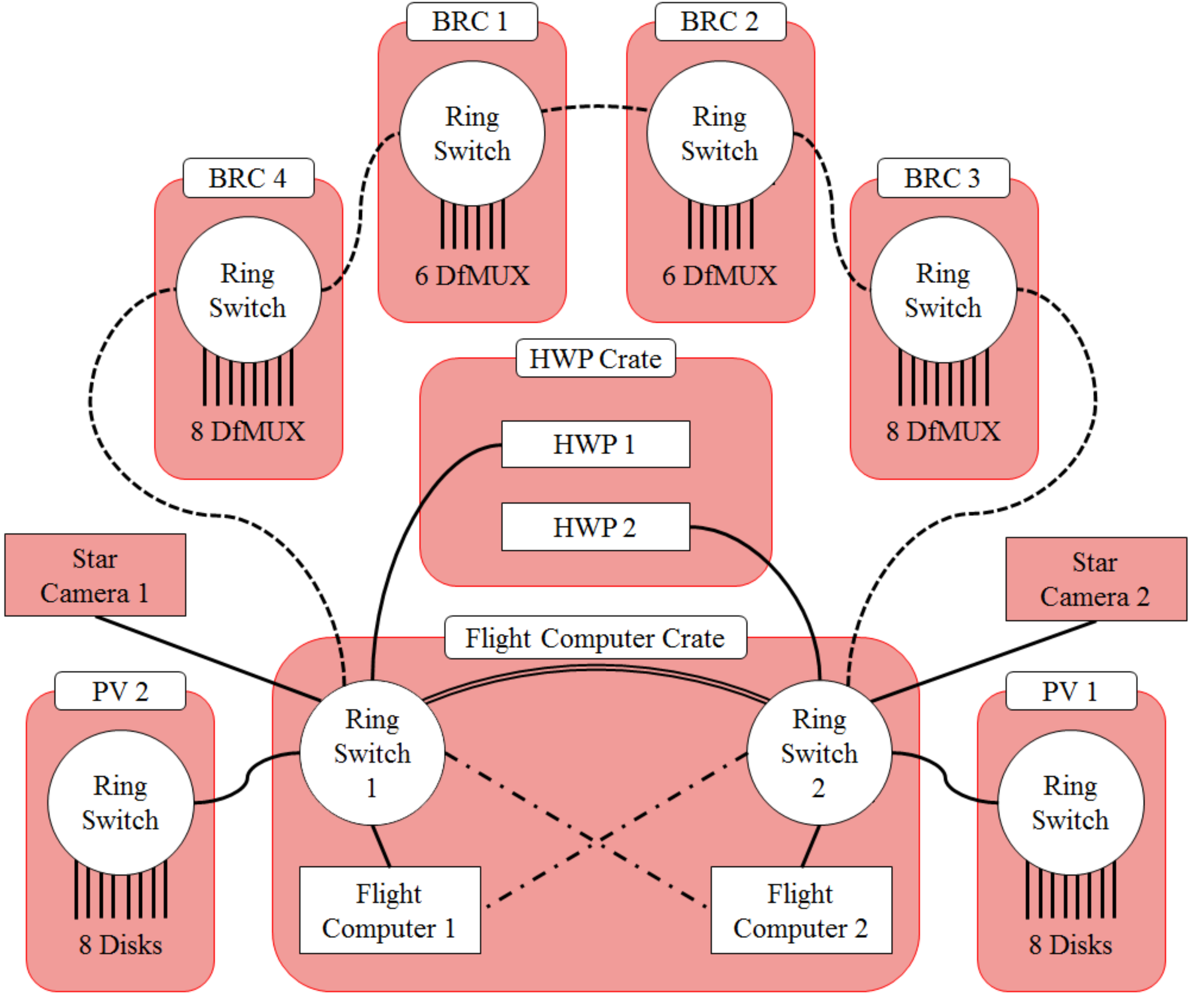}
	\caption{The \ac{EBEX} ethernet network was based on a redundant ring structure consisting of 8 ring
    switches. \ac{DfMUX} boards in each of the \ac{BRC}s communicated with a local ring switch. 
	The 4 switches were connected with fiber optic lines (dash) to
	two ring switches in the flight computer crate, which communicated with their respective flight computers via copper line connection (solid). Standard copper 
	lines also connected the flight computers with the \ac{HWP} angle readout boards and the \ac{PV} that were used to store
	data.  If any of the \ac{BRC} ring switches or fiber-optic lines malfunctioned, data from the other \ac{BRC}s would still reach the flight computers. 
	If one of the flight computer switches malfunctioned, a fail-over line activated (double line) that would pass data from the fiber optic line to the other switch, 
	and another fail-over line activated (dash-dot) connecting the non-faulty switch to both flight computers. 
	\label{fig:net-diagram} }
\end{center}
\end{figure}

We used a TCP/IP network to pass bolometer, \ac{HWP}, and star camera data to the flight computers. All 
data, including those passed to the flight computers using dedicated non-TCP/IP buses, were channeled using 
TCP/IP to two pressure vessels that held data storage disks; see Figure~\ref{fig:net-diagram}. The network 
employed a redundant ring switch technology.\footnote{Sixnet Series from Red Lion, Inc.}
Each of the 28 \ac{DfMUX} readout boards was connected via category 5e ethernet cable
to a single ring switch inside its respective \ac{BRC}. The individual ring switches were linked together 
via fiber optic lines in a redundant ring that encompassed the four readout crates and two ring switches in the 
flight computer crate.
Severing a communication link to any of the four readout crates caused 
the ring switch network to engage its backup link between the two ring switches in the 
flight computer crate. Systems that were connected to
the overall network by non-redundant links were themselves redundant with other
systems. As such, flight critical disruptions to the data network required at least two
concurrent, critical errors.

\subsection{Telemetry}
\label{sec:telemetry}

Telemetry was provided over three distinct pathways: line of sight communications,  
\ac{TDRSS}, and the Iridium satellite phone network. Line of sight communication was available only for the first
$\sim$24~hours of the flight and had a bandwidth of 1~Mbit~s$^{-1}$. Communication via the \ac{TDRSS} satellites was scheduled approximately 24~hours in advance
for several hours per day. When operational it had variable bandwidth between 6~kbit~s$^{-1}$ to 75~kbit~s$^{-1}$ and was used for both uplinking commands
and downlinking data. The Iridium phone network was available continuously with 
a bandwidth of 2~kbit~s$^{-1}$. It was used only for commanding.
Because the downlink bandwidth was variable, ground operators commanded the flight computer to throttle telemetry down to match the available bandwidth. 
The on-board throttling software relied on two elements: data compression and priority-based data downlinking.

We used a statistical data compression scheme, based on the Z-Coder \citep{Bottou_1998} adaptive binary coder.  
This compression was based on a statistical model of the bit structure of the data.  While standard statistical compression forms the 
model based only on the data available at the time of compression, the \ac{EBEX} compressor was pre-tuned to the expected form of the data streams.  
These data streams included bolometer and \ac{HWP} data, which were stored as 16-bit, big-endian timestreams; \ac{JSON} 
formatted strings generated by array tuning 
commands; text-based miscellaneous log file data; and 16-bit, little-endian housekeeping and attitude control data.
Each characteristic data stream was analyzed pre-flight and statistical models were generated and stored on the flight computer for use during flight.  
The flight control software was responsible for selecting the appropriate compression model based on the source of the data stream.  
Because the compression was still adaptive, the consequence of the flight data being different from the ground `training set' was  
merely to decrease the effective compression ratio. The compression ratio for the ground `training set' data was 10.1. Flight data showed a compression ratio of 6.2 for the bolometer data and between 7 and 10 for other 
data products. 

The compressed data were segmented into bundles of prioritized types. 
The flight software ordered these bundles by priority and fed them into the downlink stream.  Ground operators occasionally changed the the priority structure using 
standard commands so as to allow for different priorities during different phases of the flight. For example, \ac{HWP} data was prioritized when turning the 
rotation on/off, pointing information when maneuvering the attitude of the telescope, and bolometer time stream data during regular observations. 

\subsection{Tuning and Controlling the Detector Array}
\label{sec:tuningsoftware}

Operating the \ac{TES} bolometer array required tuning the bias currents in each of the 128 \ac{SQUID} 
pre-amplifiers for optimal trans-impedance~\citep{macdermid_aip2009}.  It also required
tuning the electrical bias of each of the $\sim$1000 detectors such that they were operated 
at the superconducting transition temperature. 
In practice, tuning required sending a sequential set of instructions to 
hardware components that resulted in raising and lowering 
the temperature of the \ac{SQUID}s, adjusting the 
current flowing through them, and adjusting the voltage bias of 
the bolometers. A specific set of tuning instructions is referred to as a 
`tuning algorithm.'  Each of the
28 \ac{DfMUX} electronic boards that control and readout the detectors and \ac{SQUID}s had the task of interpreting the instructions, generating the appropriate 
voltages and currents, and collecting data.   (Section~4 of \ac{EP2} provides additional  
details about the readout system and the \ac{DfMUX} boards.) 

In a typical ground-based experiment, a central control computer cycles through each step in the algorithms for each of the 
\ac{SQUID}s and \ac{TES}s, sends commands to modify the configuration of the readout boards, and collects and stores the data.
Because of the limited uplink bandwidth we developed 
an efficient way to address each of the \ac{SQUID}s and detectors in the array. 
Because of the limited computing power available we transferred execution of the tuning algorithms from the 
flight computers to each of the 28 electronic boards controlling the \ac{SQUID}s and \ac{TES} detectors. Finally, limited downlink bandwidth required that we 
economize diagnostic data sent to the ground as a result of tuning the array. We now describe each of these developments. 

\subsubsection{Addressing SQUID and TES Commands} 
\label{sec:squid_tes_commands}

To efficiently address an individual \ac{TES} or \ac{SQUID}, we constructed a 
`hardware map' that stored the mapping between each \ac{TES}, its readout \ac{SQUID}, 
its \ac{DfMUX} board, and all other readout components. It also stored the 
pre-selected parameters to setup the entire array, determined during pre-flight 
testing.

The hardware map was stored as a series of linked tables in an sqlite3 
database~\citep{Newman_2004} on the internal disk on each of the two flight computers.  Tuning commands were then 
targeted to any arbitrary subsection of the hardware by forming a request using \ac{SQL} 
commands. Such commands were generated automatically on-board the 
flight computer during pre-scheduled tunings of the array  
or, as necessary, by ground operators.

The \ac{SQL} commands had the advantage that they were sufficiently compact 
as to fit in the 250-byte maximum command length set by the \ac{CSBF} uplink protocol
and still flexible enough to provide ground operators full manual tuning capability. For example, 
after the first tuning at float we discovered that a significant fraction of 
detector biases, over 70\%, needed to be adjusted. This was anticipated 
pre-flight as we lacked a sufficiently accurate estimate
of the millimeter-wave emission of the atmosphere. The \ac{SQL} commands enabled flexible and relatively 
rapid retuning of the array and storage of new default parameters for each 
detector. 

\subsubsection{Executing the Tuning Algorithms}

We transferred the responsibility of executing the tuning algorithms from a central computer to 
the \ac{DfMUX} boards. A library of tuning algorithms written in Python was loaded 
onto a flash memory card installed on each of the \ac{DfMUX} boards. 
The commands arriving from the on-board computer consisted of a single instruction for an entire tuning 
algorithm, instead of a line-by-line cycle through the algorithm itself. When such a command arrived, the resident 
MicroBlaze soft processor running on the \ac{DfMUX} board's   FPGA\footnote{Virtex-4 FPGA by Xilinx Inc.} 
executed the algorithm \citep{MacDermid_masters_thesis}. An `algorithm-manager' program 
running on the MicroBlaze delegated the tuning to several processes that ran as 
parallel threads. This allowed multiple \ac{TES} bolometers connected to the same \ac{DfMUX} 
board to be tuned simultaneously, saving tuning time. Tuning algorithms that 
were executed on different boards were always executed in parallel. Therefore total array 
tuning time was independent of the size of the array. 

\subsubsection{Handling Diagnostic Data}

Each tuning algorithm collected diagnostic data on the components it tuned, such as the $I-V$ characteristics 
of each bolometer as it was biased into its superconducting transition.
The entirety of the tuning data were stored on-board as they were vital for understanding the performance of the 
bolometer array during post-flight analysis.
The data were also important for monitoring and tuning 
the bolometer array during flight, but sufficient telemetry bandwidth was not always assured. We therefore 
used the telemetry prioritization scheme described in Section~\ref{sec:telemetry}. After
each tuning algorithm completed execution we first downlinked the current state of the \ac{DfMUX} boards' settings. 
Ground operators compared the current end-of-algorithm state 
of the boards to the initial pre-algorithm execution state to ascertain that the algorithm was in fact executed. 
We then prioritized downlinking the entire data available for each of the tuning algorithms. These data had the highest
priority while any array tuning activities took place as there were no other scientifically interesting data available during 
that time. To quickly digest and act upon the influx of tuning data we wrote custom software that automatically analyzed them 
and made the results available in automatically generated web pages~\citep{MacDermid_thesis}. 
Ground operators used the results to plan the subsequent actions required to optimize the performance of the bolometer array.

In many cases some of the tuning data were not fully downlinked by the time we resumed science observations, at 
which time attitude control and detector time stream data received higher telemetry priority. Array tuning data
that received lower downlink priority typically trickled down over several hours after the completion of the tuning 
operation. 

\subsection{Data Management}

We designed the data management and storage system to handle all the data that could be collected by the 1792~readout channels available 
with the 28~\ac{DfMUX} electronic readout boards. With 16 bits per sample and a sampling rate of 190.735~Hz the anticipated data rate was 5.5~Mbits~s$^{-1}$
accumulating to 590~GB over 10~days of flight. At the time we designed the system, this volume exceeded the capacity of any single, 
commercially available hard drive. We therefore designed two redundant disk arrays based on 
\ac{ATA} over Ethernet (AoE) protocol.  
Each array contained a full copy of all flight data including science data, system logs, attitude sensor data, and housekeeping information. 
All data were written to disk in a standardized packet format. All packets had a header with information that included 
data source identification -- for example, bolometer identifier or temperature sensor identifier -- and a time stamp.

Each array had seven 320~GB commercial 2.5"~ parallel \ac{ATA} magnetic hard disks. 
Three disks were sufficient to hold a complete copy of the pre-flight and flight data, with ample margin; the other disks 
were designed to remain empty and be used only 
in case of disk hardware failure. Each array was housed in a separate pressure vessel in which we maintained atmospheric pressure and 
circulated the air with two fans. (We decided to not use solid state memory due to cost and concerns about the robustness of this hardware under the elevated cosmic ray flux at float altitude.) 
During pre-flight integration and testing, vibrations from the fans initially induced repeated disk failures. 
Decoupling the fan mounting from the structure holding the disks eliminated disk failures. 
Following this modification, we did not experience any further disk failure throughout integration of the instrument or flight.    
The hard drives were mounted on printed circuit boards called Blade~II that implemented the AoE protocol.\footnote{Coraid, Inc.}
We found that, with the Blade~II, the rate at which data were written to disk declined approximately linearly with an increase 
in the number of files being written simultaneously. We combined similar data streams into the same files resulting 
in a maximum of 31 files open at any given time. At this level the writing rate was adequate and no data were lost due to pile-ups. 

We developed custom disk management software that wrote data to a single disk within the array to minimize power consumption by keeping 
all other disks idle.  Steady state power consumption was measured to be 38.4~W per disk pressure vessel. 
The software maintained information about the current state of each disk including its error count, free space, and response speed.
Using these data, the software selected the best available disk on startup and wrote all data to the same disk until it either filled or failed.
The software then selected the next available disk, mounted the new disk in a background process, and moved data streaming to the new disk.
During the transition, which was measured to take 5 s, data were buffered; buffer depth allowed for 75 s of data storage.


\section{Summary}
\label{sec:summary}

To probe the CMB {\it E}- and {\it B}-modes with higher sensitivity compared to previous instruments 
and at frequency bands not accessible for ground observatories, we built EBEX, a stratospheric  
balloon-borne instrument with approximately 1000 detectors and designed for long-duration Antarctic flights.

EBEX pioneered the use of TES bolometers on a balloon-borne platform. It was the first experiment to fly 
a small array of these detectors in a test-flight in 2009 and 
a kilo-pixel array during its EBEX2013 flight. Nearly 1000 TES bolometers 
were operating shortly after the payload reached float altitude, and subsequent refrigerator cycles and array tuning 
operations could proceed with little ground-operator intervention.

The platform presented unique challenges in computing power, 
bandwidth, and duration of observation time. To meet these challenges we developed a flexible flight event scheduler, an on-board network, 
a tuning and control software based on SQL commands, a prioritized telemetry downlink, and an efficient method to implement 
all array tuning commands. 

To limit the number of wires reaching the lowest temperature cryogenic stage, we decided to multiplex the readout of 
several detectors onto a single pair of bias and readout wires. We chose frequency domain multiplexing~\citep{dobbs_revSciInst_2012} -- as opposed to time 
domain multiplexing~\citep{ACTpolTDM} -- because it required fewer wires at the lowest temperature stage, simplifying the design 
of the cryogenic receiver.  This decision moved the design and implementation challenge from optimizing the cryogenic stage to two 
other aspects: (1) developing a frequency domain multiplexing system that had lower power consumption compared to the system available 
at the time the project began, and (2) providing and dissipating the still-appreciable level of heat generated by the new system. We pioneered 
the implementation of the \ac{DfMUX} system, now used by a number of other experiments; this is discussed more fully in \ac{EP2}. 
We described the EBEX power system that could provide more than 2.3~kW for a calculated peak 
load of 1.7~kW, of which 590~W was consumed by the readout system.
Much of the power dissipation was localized in few electronic components. Conduction, heat pipes, and a liquid cooling loop were used 
to transfer the energy to panels that radiated it to the sky. In-flight performance matched pre-flight predictions and showed that all electronic 
boards stayed within their nominal operating temperatures.

The optical system including both telescope and receiver, which was sized to accommodate
the kilo-pixel array and to reach 6\am~resolution, led to an 8-m tall gondola and a payload weight of 2810~kg that together
with NASA equipment approached the balloon load limit of 3600~kg. We gave an overview of the mechanical structure 
of the \ac{EBEX} gondola. A number of measures to reduce weight are described throughout the series of three papers. 
One of them was the use of polyethylene suspension cables as 
part of the gondola structure. To our knowledge, this is the first use of such cables in a stratospheric, long-duration balloon payload. 
The EBEX2013 experience indicates that these cables are suitable for balloon flights as long as proper consideration is given to 
their UV sensitivity and initial creep. 

Telescope elevation motion was achieved by moving the inner relative to the outer frame with a linear actuator. 
We described a mechanism to lock the inner relative to the outer frame and thus reduce risk of damage to the linear
actuator from launch accelerations. We controlled azimuth motion with a pivot and a reaction wheel. 
A thermal model error caused the pivot motor controller to overheat and turn-off when attempting to 
execute the design scan strategy. Free gondola azimuth motion consisted of a superposition of full rotations 
at variable rotational speed and 
an 80~s period oscillatory motion at the natural torsional period of the flight line and gondola. We therefore
chose to fix the elevation, giving a sky scan consisting of a 5700~square degree strip in DEC. 

We described the EBEX2013 attitude determination system, which relied on 
star cameras, gyroscopes, a custom-built star camera software STARS, and an attitude reconstruction software. 
The in-flight sky scan, with nearly zero azimuth speed occurring approximately every 40~s, matched the pre-flight plans
for attitude reconstruction. The reconstruction software was used to minimize attitude errors and to assess their contribution to spurious {\it B}-modes.  
The combination of hardware, STARS, and the attitude reconstruction software constrained 
attitude errors such that the spurious {\it B}-modes they induced are less than 10\% of predicted CMB {\it B}-modes with $r=0.05$ for $30 \lesssim \ell \lesssim 1500$. 

Two companion papers provide additional details about the EBEX instrument's telescope, receiver, and 
polarimetry (\ac{EP1}) and the detectors, their readout, and their flight performance (\ac{EP2}).

\begin{acronym}
    \acro{ACS}{attitude control system}
    \acro{ADC}{analog-to-digital converter}
    \acrodefplural{ADC}[ADCs]{analog-to-digital converters}
    \acro{ADS}{attitude determination software}
    \acro{AMC}{Advanced Motion Controls}
    \acro{ARC}{anti-reflection coating}
    \acro{ATA}{Advanced Technology Attachment}
    \acrodefplural{BRC}[BRCs]{bolometer readout crates}
    \acro{BRC}{bolometer readout crate}
    \acro{BLAST}{Balloon-borne Large-Aperture Submillimeter Telescope}
    \acro{CAN bus}{Controller Area Network bus}
    \acro{CMB}{cosmic microwave background}
    \acro{CMM}{coordinate measurement machine}
    \acro{CSBF}{Columbia Scientific Balloon Facility}
    \acro{CCD}{charge coupled device}
    \acro{DAC}{digital-to-analog converter}
    \acrodefplural{DAC}[DACs]{digital-to-analog converters}
    \acro{DASI}{Degree~Angular~Scale~Interferometer}
    \acro{dGPS}{differential global positioning system}
    \acro{DfMUX}{digital~frequency~domain~multiplexer}
    \acro{DLFOV}{diffraction limited field of view}
    \acro{DSP}{digital signal processing}
    \acro{EBEX}{E~and~B~Experiment}
    \acro{EBEX2013}{EBEX2013}
    \acro{ELIS}{EBEX low inductance striplines}
    \acro{EP1}{EBEX Paper 1}
    \acro{EP2}{EBEX Paper 2}
    \acro{EP3}{EBEX Paper 3}
    \acro{ETC}{EBEX test cryostat}
    \acro{FDM}{frequency domain multiplexing}
    \acro{FPGA}{field programmable gate array}
    \acro{FCP}{flight control program}
    \acro{FOV}{field of view}
    \acro{FWHM}{full width half maximum}
    \acro{GPS}{global positioning system}
    \acro{HPE}{high-pass edge}
    \acro{HWP}{half-wave plate}
    \acro{IA}{integrated attitude}
    \acro{IP}{instrumental polarization} 
    \acro{JSON}{JavaScript Object Notation}
    \acro{LDB}{long duration balloon}
    \acro{LED}{light emitting diode}
    \acro{LCS}{liquid cooling system}
    \acro{LC}{inductor and capacitor}
    \acro{LPE}{low-pass edge}
    \acro{MLR}{multilayer reflective}
    \acro{MAXIMA}{Millimeter~Anisotropy~eXperiment~IMaging~Array}
    \acro{NASA}{National Aeronautics and Space Administration}
    \acrodefplural{NASA}[NASA's]{National Aeronautics and Space Administration's}
    \acro{NDF}{neutral density filter}
    \acro{PCB}{printed circuit board}
    \acro{PE}{polyethylene}
    \acro{PTFE}{polytetrafluoroethylene}
    \acro{PME}{polarization modulation efficiency}
    \acro{PSF}{point spread function}
    \acro{PV}{pressure vessel}
    \acro{PWM}{pulse-width modulated}
    \acro{RMS}{root mean square}
    \acro{SLR}{single layer reflective}
    \acro{SMB}{superconducting magnetic bearing}
    \acro{SQUID}{superconducting quantum interference device}
    \acro{SQL}{structured query language}
    \acro{STARS}{Star Tracking Attitude Reconstruction Software}
    \acro{TES}{transition edge sensor}
    \acro{TDRSS}{tracking and data relay satellites}
   \acro{TM}{transformation matrix}
   \acro{UTC}{Coordinated Universal Time}

\end{acronym}

\acknowledgments
Support for the development and flight of the EBEX instrument was provided by NASA grants NNX12AD50G, NNX13AE49G, NNX08AG40G, 
and NNG05GE62G, and by NSF grants AST-0705134 and ANT-0944513.   
Ade, Moncelsi, Pascale, and Tucker acknowledge the 
Science \& Technology Facilities Council for its continued support of the underpinning technology for filter and waveplate development.  
We also acknowledge support by the Canada Space Agency, the Canada Research Chairs Program, the Natural Sciences and 
Engineering Research Council of Canada, the Canadian Institute for Advanced Research, 
the Minnesota Supercomputing Institute at the University of Minnesota, 
the National Energy Research Scientific Computing Center, the Minnesota and Rhode Island 
Space Grant Consortia, our collaborating institutions, and Sigma Xi the Scientific Research Society. 
Baccigalupi acknowledges support by the RADIOFOREGROUNDS project, funded by the European Commission’s H2020 Research Infrastructures under the Grant Agreement 687312, as well as the Italian INFN INDARK initiative.
Didier acknowledges a NASA NESSF fellowship NNX11AL15H.  Helson acknowledges NASA NSTRF fellowship NNX11AN35H.  
Reichborn-Kjennerud acknowledges an NSF Post-Doctoral Fellowship AST-1102774, and a NASA Graduate Student Research Fellowship. Raach and Zilic acknowledge support by the Minnesota Space Grant Consortium.  
We very much thank Danny Ball and his colleagues at the Columbia Scientific Balloon Facility for their dedicated support of the EBEX program.  
We thankfully acknowledge contributions to predicting systematic errors by Matias Zaldarriaga and Amit Yadav.  We thank the BLAST team 
for providing the original version of the flight computer program discussed in Section~\ref{sec:flightmanagement}. 
We thank Christopher Geach, Qi Wen, and Xin Zhi Tan for help with figures.

\bibliographystyle{aasjournal}
\bibliography{EBEXPaper3}{}

\end{document}